\documentclass[=journal]{IEEEtran}
\usepackage{amsmath,amsfonts,amssymb}
\usepackage{algorithmic}
\usepackage{algorithm}
\usepackage{array}
\usepackage[caption=false,font=normalsize,labelfont=sf,textfont=sf]{subfig}
\usepackage{textcomp}
\usepackage{stfloats}
\usepackage{verbatim}
\usepackage{graphicx}
\usepackage{cite}
\usepackage{multirow}
\usepackage[normalem]{ulem}
\usepackage{newtxtext}
\usepackage{newtxmath} 
\let\mathcal\undefined
\DeclareMathAlphabet{\mathcal}{OMS}{cmsy}{m}{n}
\usepackage{bm}

\usepackage{amssymb}
\usepackage{utfsym}
\useunder{\uline}{\ul}{}
\usepackage{graphicx}
\usepackage{float}
\usepackage[colorlinks,urlcolor=blue,linkcolor=blue,citecolor=blue]{hyperref}
\usepackage{cleveref}
\usepackage{fancyhdr}

\fancypagestyle{firstpage}{
    \fancyhf{}
    \fancyhead[R]{\thepage}
    \fancyfoot[C]{This work has been submitted to the IEEE for possible publication. Copyright may be transferred without notice, after which this version may no longer be accessible.}
}

\begin{document}

\title{Cross-Domain Separable Translation Network for Multimodal Image Change Detection}

\author{~Tao~Zhan,~\IEEEmembership{Member,~IEEE},~Yuanyuan~Zhu,~Jie~Lan,~and~Qianlong~Dang
\thanks{This work was supported in part by the National Natural Science Foundation of China under Grant 62103311, in part by the Chinese Universities Scientific Fund under Grant 2452024408, in part by the Natural Science Foundation of Shaanxi under Grant 2024JC-YBQN-0687, and in part by the Fundamental Research Funds of the Central Universities of China under Grant 2452023013. (\emph{Corresponding author: Tao Zhan.})}

\thanks{Tao Zhan, Yuanyuan Zhu and Jie Lan are with the Shaanxi Engineering Research Center of Agricultural Information Intelligent Perception and Analysis, College of Information Engineering, Northwest A\&F University, Yangling, Xianyang 712100, China (e-mail: omegazhant@gmail.com).}
\thanks{Qianlong Dang is with the College of Science, Northwest A\&F University, Yangling, Xianyang 712100, China.}
}



\maketitle
\thispagestyle{firstpage}

\begin{abstract}
In the remote sensing community, multimodal change detection (MCD) is particularly critical due to its ability to track changes across different imaging conditions and sensor types, making it highly applicable to a wide range of real-world scenarios. This paper focuses on addressing the challenges of MCD, especially the difficulty in comparing images from different sensors with varying styles and statistical characteristics of geospatial objects. Traditional MCD methods often struggle with these variations, leading to inaccurate and unreliable results. To overcome these limitations, a novel unsupervised cross-domain separable translation network (CSTN) is proposed, which uniquely integrates a within-domain self-reconstruction and a cross-domain image translation and cycle-reconstruction workflow with change detection constraints. The model is optimized by implementing both the tasks of image translation and MCD simultaneously, thereby guaranteeing the comparability of learned features from multimodal images. Specifically, a simple yet efficient dual-branch convolutional architecture is employed to separate the content and style information of multimodal images. This process generates a style-independent content-comparable feature space, which is crucial for achieving accurate change detection even in the presence of significant sensor variations. Extensive experimental results demonstrate the effectiveness of the proposed method, showing remarkable improvements over state-of-the-art approaches in terms of accuracy and efficacy for MCD. The implementation of our method will be publicly available at \url{https://github.com/OMEGA-RS/CSTN}.
\end{abstract}


\section{Introduction}
\subsection{Background}
\IEEEPARstart{T}{he} concept of change detection in remote sensing involves identifying changes in an object's or phenomenon's state through observations made at distinct times \cite{singh1989review}. Advancements in remote sensing technology have markedly enhanced both the convenience and quality of acquiring remote sensing data \cite{cheng2024change}. These data encompass various types of images, such as common optical RGB images, multispectral/hyperspectral images, synthetic aperture radar (SAR) images, and LiDAR point cloud images, each offering unique advantages for different applications. As a fundamental task in remote sensing community, change detection has consequently been the focus of extensive research and application across multiple domains, including land cover monitoring \cite{mohamed2020monitoring}, \cite{zhang2024glc_fcs30d}, urban planning\cite{han2010change}, \cite{stilla2023change}, agricultural surveys\cite{gim2020improved}, \cite{liu2022cnn}, and disaster assessment\cite{brunner2010change}, \cite{tsvetkov2023change}, \cite{yin2023research}.

Change detection technology can be broadly classified into two categories based on the timestamps of acquired images: bi-temporal change detection and multi-temporal change detection \cite{jianya2008review}. Bi-temporal change detection focuses on identifying changes between two specific dates, primarily for assessing immediate impacts such as those caused by natural disasters \cite{janalipour2017building}, \cite{lu2019landslide}. Multi-temporal change detection, or time series change analysis, examines changes over a continuous time scale, making it suitable for monitoring long-term trends \cite{lunetta2006land}, \cite{lu2016spatio}. While bi-temporal methods are often employed for rapid assessment, multi-temporal approaches provide deeper insights into evolving patterns and processes. Considering that multi-temporal change detection can be decomposed into bi-temporal change detection and long-term serial analysis \cite{jianya2008review}, many current studies are focused on bi-temporal data due to its wide range of applications.

The categorization of change detection techniques based on data sources results in two main categories: unimodal change detection (UCD), also known as homogeneous change detection, and multimodal change detection (MCD), also referred to as heterogeneous change detection \cite{lv2022land}, \cite{chen2022unsupervised}, \cite{sun2022iterative}. UCD, which utilizes identical sensors and imaging principles, is an intuitive approach that has garnered significant attention \cite{lv2022land}, \cite{hemati2021systematic}, \cite{lv2021land}, \cite{wen2021change}, \cite{lv2022spatial}, \cite{li2020deep}, \cite{wang2018getnet}. However, its applicability is limited in scenarios where unimodal images are unavailable, such as during sudden-onset disasters when pre-event optical images are available but post-event optical images are hindered by adverse weather conditions. In contrast, MCD, which leverages images from disparate sensors, can effectively bridge this gap \cite{brunner2010earthquake}, \cite{prendes2014new}, \cite{touati2017energy}. Despite exhibiting variations in style and statistical behavior, these multimodal images can still be processed to achieve satisfactory change detection results, making MCD a vital approach in certain situations \cite{lv2021land}.

\subsection{Related Work}
\label{RelatedWork}
The discrepancies in appearances and statistical characteristics between multimodal images present a significant challenge in MCD: the inability to directly compare these images for change detection \cite{shafique2022deep}. To address this issue, numerous methods have been proposed to transform incomparable multimodal images into a comparable space. Drawing upon the taxonomy presented in \cite{sun2021iterative}, but with certain refinements and extensions, this study classifies MCD methods into four categories based on the spatial domain in which changes are identified: 1) classification-based methods; 2) handcrafted feature-based methods; 3) deep latent feature-based methods; and 4) image translation-based methods. This subsection will further elaborate on each of these categories.

\subsubsection{Classification-Based Methods}
Methods in this category typically operate in the land-cover classification space, where images from disparate time periods were classified at either the pixel-based \cite{serra2003post}, \cite{colditz2012potential}, \cite{li2021spatially} or object-based level \cite{qin2013object}, \cite{wan2019object}. Through a comparative analysis of the resulting classification maps, changes can be effectively identified. A notable advantage of these methods lies in their ability to directly capture land-cover type changes, yielding ``from-to" change maps that facilitate in-depth analysis and interpretation. However, the accuracy of these methods is contingent upon the performance of the classification algorithms and the quality of the training samples \cite{yuan2005land}. Furthermore, the granularity of change detection is also limited by the resolution of the classification, as coarse classification schemes may fail to capture fine-grained changes \cite{sun2021iterative}. Notably, classification errors can propagate through the process, potentially impacting the reliability of the change detection results.

\subsubsection{Handcrafted Feature-Based Methods}
These methods rely on handcrafted feature representations of basic units (e.g., pixels, superpixels, patches) designed based on domain knowledge or specific assumptions. Change detection is performed by comparing the similarity of corresponding units in the pre-event and post-event images within this engineered feature space. For example, Wan et al.\cite{wan2018multi} used sorted histograms to represent image patches, and computed distances between patches to detect changes. Touati et al. \cite{touati2018change} proposed a modality-invariant multidimensional scaling representation, which is used to measure differences between multimodal images. Sun et al. \cite{sun2021iterative} introduced an iterative robust graph and Markovian co-segmentation method (IRG-McS) which constructs a $K$-nearest neighbor graph of superpixels, where edge weights represent distances between nodes. This graph was then compared to the corresponding superpixels in the target domain to evaluate the extent of their dissimilarity. Methods like NLPG\cite{sun2021nonlocal} and its improved version INLPG\cite{sun2021structure} were developed by building a nonlocal patch similarity-based graph to measure structural differences between patch pairs. These methods are subject to manually defined representations, which may not accurately capture differences between multimodal images, leading to inaccurate change detection results.

\subsubsection{Deep Latent Feature-Based Methods}
This category of methods utilizes deep neural networks to transform multimodal images into a comparable feature space, where change detection can be performed. A symmetric convolutional coupling network (SCCN) \cite{liu2016deep} was proposed to highlight changes by shrinking the feature differences in unchanged regions through the transformation of multimodal images into a consistent feature space. Zhan et al. introduced the iterative feature mapping network (IFMN)\cite{zhan2018iterative}, which extracts high-level features from multi-source images and aligns them through an iterative process to enable change detection. Other approaches, such as the probabilistic model based on bipartite convolution network (PMBCN)\cite{liu2021probabilistic} and the commonality autoencoder change detection (CACD)\cite{wu2021commonality} method, were also developed to leverage deep learning for capturing feature distributions and exploring commonalities between images, respectively. These methods have demonstrated exceptional performance in MCD tasks. The key to their performance lies in designing suitable network architectures and workflows to enhance feature representation quality, as well as formulating appropriate objective functions to constrain the learning of deep neural networks.

\subsubsection{Image Translation-Based Methods}
This category of methods leverages image translation techniques to directly translate multimodal images into the corresponding modality of the other. This direct translation then allows for change detection in the image space. Image translation is a computer vision task that seeks to convert an image from a source domain to a target domain while maintaining the inherent semantic content from the source and incorporating the desired stylistic properties of the target\cite{Pang2021}. Researchers have widely embraced image translation as a powerful tool for MCD. The homogeneous pixel transformation (HPT) method \cite{liu2017change} was proposed to translate images by creating pixel-wise correspondences between the original and target feature spaces, facilitating accurate change detection. Niu et al. \cite{niu2018conditional} devised the conditional adversarial network (CAN), which bridges the gap between SAR and optical images for MCD by using a conditional generative adversarial network and an approximation network. X-Net and ACE-Net were introduced by Luppino et al. \cite{luppino2021deep}, which demonstrate competitive performance in unsupervised MCD by leveraging affinity-based change priors and weighted loss functions, highlighting their effectiveness in image translation and change detection. Luppino et al. \cite{luppino2022code} also proposed code-aligned autoencoders (CAAE), which utilize image translation and domain-specific affinity matrices to align code spaces of two autoencoders, effectively minimizing the contribution of change pixels to the learning process and performing change detection in the image domain. The adaptive graph-based structure consistency (AGSCC) model \cite{sun2022image} pioneered a structure consistency-based image regression approach for image translation, utilizing an adaptively learned graph and regularization terms to preserve structure in the transformed image. Sun et al. \cite{sun2023structural} proposed the structural regression fusion (SRF) method, which addresses the challenge of structural asymmetry in MCD by constructing a hypergraph for image translation, yielding promising results. Li et al. \cite{li2024comic} presented a copula mixture and CycleGAN-based change detection method (COMIC) to achieve robust unsupervised MCD. Most of these methods typically treat image translation and change detection as distinct tasks. Consequently, the accuracy of change detection is directly influenced by the output of the image translation process.

\begin{figure*}[htbp]
	\centering
	\includegraphics[width=1\textwidth]{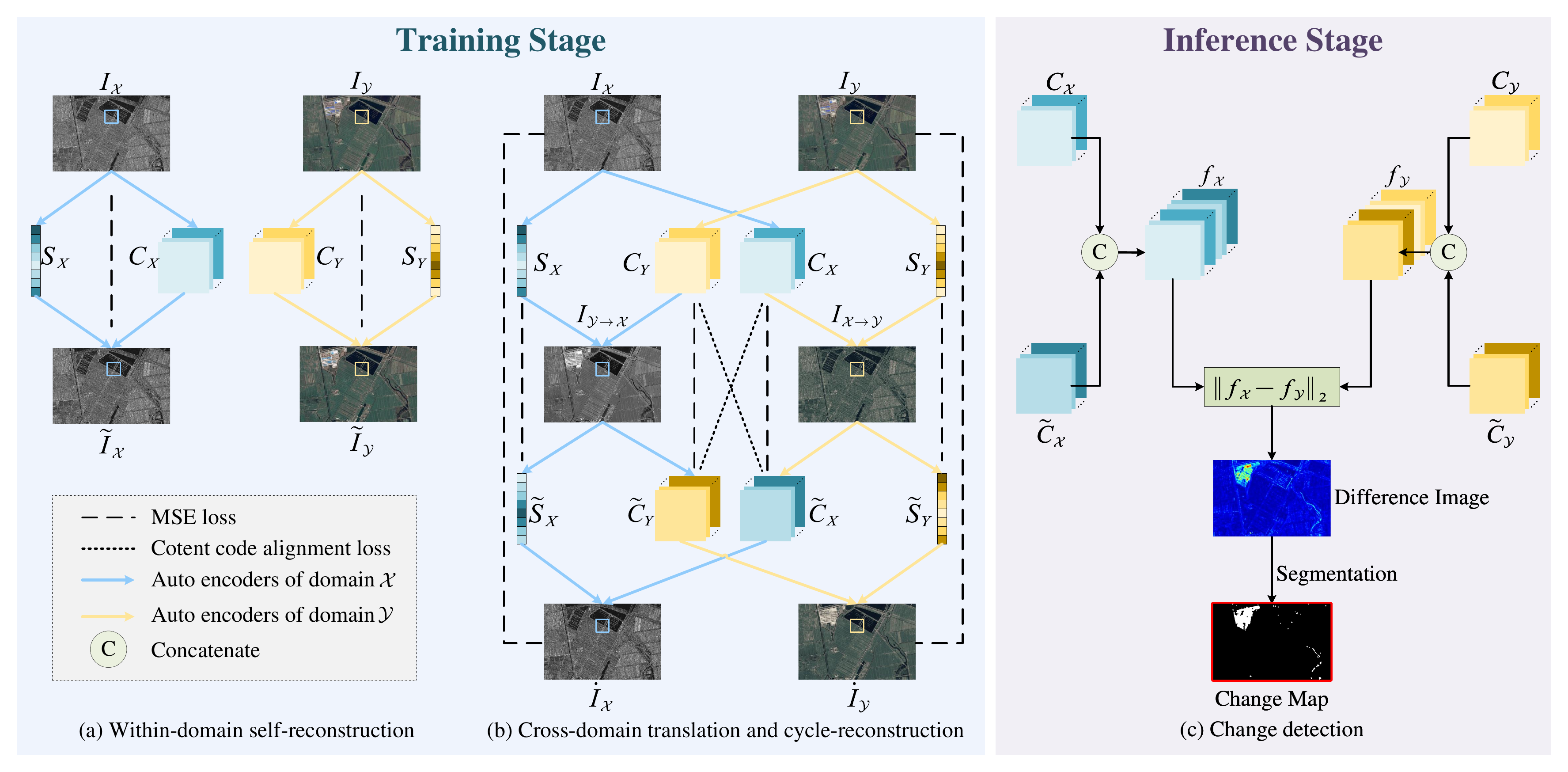}
	\caption{Overview of the proposed CSTN framework. During the training stage, the model is optimized by jointly executing following two workflows: (a) within-domain image self-reconstruction and (b) cross-domain image translation and cycle-reconstruction, aiming to acquire comparable content features from multimodal images by establishing a style-independent feature space. In the inference stage, (c) the change detection pipeline is performed to derive a binary change map by making full use of the content features from the multimodal mages and translated images.}
	\label{fig1}
\end{figure*}
\subsection{Motivation}
Driven by advancements in deep learning, deep latent feature-based methods and image translation-based methods show greater promise. However, these methods also face significant challenges. 

Deep latent feature-based methods require carefully designed model architectures and objective functions tailored to the specific requirements of MCD, leading to increased complexity in optimization and implementation. Such methods, while powerful, can be sensitive to the choice of hyperparameters and may suffer from overfitting, which limits their practical usability. Furthermore, their performance can vary significantly depending on the quality and variability of the input data. Further exploration is crucial to address these complexities and improve the robustness and generalization capabilities of this kind of methods.

Image translation-based methods also have inherent flaws, primarily because treating image translation and MCD as separate stages can lead to translation errors that compromise the final MCD results. The quality of image translation might not meet the requirements of MCD, as the process can introduce alterations or obscure subtle changes crucial for accurate detection. Moreover, the use of generative adversarial networks (GANs) adds adversarial loss, escalating computational complexity. Furthermore, the necessity for careful tuning of loss weights may hinder generalization across various datasets. These methods typically rely solely on image comparison for detecting changes, potentially overlooking the rich information embedded in intermediate features. Future research should prioritize the development of methods that tightly integrate translation and change detection tasks. 

Given the aforementioned analysis, we draw inspiration from both deep latent feature-based and image translation-based approaches to advance deep feature extraction frameworks. We present a novel cross-domain separable translation network (CSTN), which employs a within-domain self-reconstruction and a cross-domain image translation and cycle-reconstruction workflow to decompose images into content and style codes, facilitating style-independent content codes for change detection.

\subsection{Contribution}
The contributions of this work are summarized as follows:
\begin{enumerate}
    \item We propose a novel multi-task workflow with change detection constraint to separate content and style representations, creating a style-independent feature space inherently suitable for accurate change detection. 
    \item We present a unified framework that seamlessly integrates image translation and MCD, enabling joint optimization for enhanced performance for both tasks.
    \item We simplify the training process by employing equal weights for loss components in CSTN, eliminating the need for hyperparameter tuning, thereby enhancing usability and reducing implementation complexity.
    \item Extensive experiments demonstrate that CSTN excels in both MCD and image translation, showcasing its versatility and effectiveness in addressing issues posed by deep latent feature-based and image translation-based methods for MCD.
\end{enumerate}

\subsection{Outline}
The rest of this article is structured as follows. Section \ref{Section:method} describes the proposed CSTN in detail. Section \ref{Section:experiments} presents the experimental analysis and discussion. The conclusion of this work is provided in Section \ref{Section:Conclusion}.

\section{Methodology}
\label{Section:method}
\subsection{Overview}
This study aims at identifying land surface changes between two co-registered images acquired by disparate sensors. The inherent differences in sensor modalities lead to distinct characteristics in image style, resulting in different feature spaces associated with domains. For instance, a SAR image resides in the gray-scale space, while an optical image occupies the spectral space. This disparity hinders direct comparison for change detection. Specifically, $I_\mathcal{X}\in \mathbb{R} ^{H\times W\times C_1}$ and $I_\mathcal{Y}\in \mathbb{R} ^{H\times W\times C_2}$ represent a pre-event image and a post-event image in domains $\mathcal{X}$ and $\mathcal{Y}$, respectively, where $H$ and $W$ denote the height and width of the images, $C_1$ and $C_2$ denote the corresponding channels. Both images are divided into image patches denoted as $X$ and $Y$, where $X \in \mathbb{R}^{h \times w \times C_1}$ and $Y \in \mathbb{R}^{h \times w \times C_2}$, with $h$ and $w$ representing the height and width of each patch.

Inspired by multimodal unsupervised image-to-image translation (MUNIT) \cite{huang2018multimodal}, we propose a novel framework utilizing a dual-path approach for training our MCD model. This framework incorporates a within-domain image self-reconstruction and a cross-domain image translation and cycle-reconstruction workflow. In the within-domain workflow, images $I_{\mathcal{X}}$ and $I_{\mathcal{Y}}$ are encoded and directly reconstructed as $\tilde{I}_{\mathcal{X}}$ and $\tilde{I}_{\mathcal{Y}}$. This process ensures the encoder can effectively extract both content and style information, and the decoder can reconstruct the images using given information. In the cross-domain workflow, for images $I_{\mathcal{X}}$ and $I_{\mathcal{Y}}$, we first encode each image to obtain their respective content and style codes. These codes are then recombined across domains to generate translated images $I_{\mathcal{X} \rightarrow \mathcal{Y}}$ and $I_{\mathcal{Y} \rightarrow \mathcal{X}}$. Subsequently, a second encoding of these translated images is performed, followed by reconstruction to produce the final outputs $\dot{I}_{\mathcal{X}}$ and $\dot{I}_{\mathcal{Y}}$. Several loss components, computed using the elements derived from the two workflows mentioned above, are designed to guide the optimization of our model. To be more specific, the model parameters are optimized to develop encoders capable of encoding images into content and style, as well as decoders that can perform either image translation or reconstruction based on different inputs. This approach allows for the extraction of style-independent content representations and thus achieves MCD. A detailed description of the loss functions is provided in Section \ref{subsection:lossfunction}. The overall flowchart of the proposed model is illustrated in Fig. \ref{fig1}. 
\begin{figure}[tbp!]
	\centering
	\includegraphics[width=0.45\textwidth]{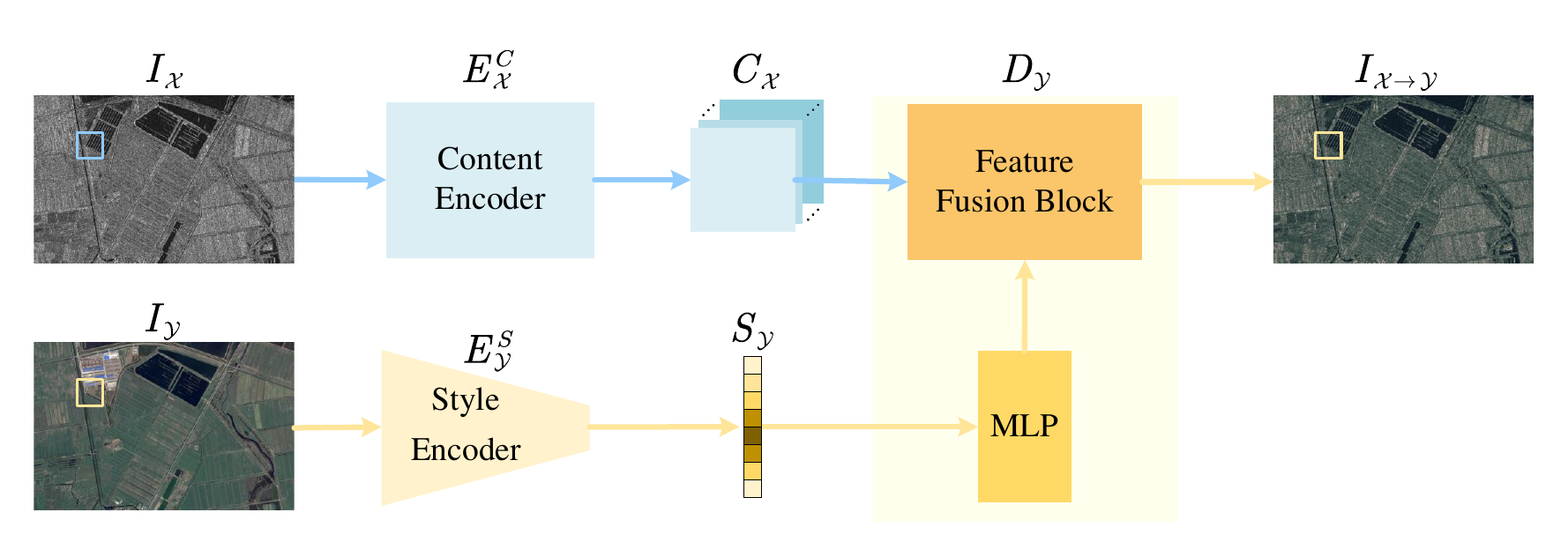}
	\caption{Scheme of cross-domain image translation (taking domain $\mathcal{X}$→ domain $\mathcal{Y}$ as an example).}
	\label{fig2}
\end{figure}
\subsection{Model Architecture}
The proposed model introduces a novel approach to image manipulation by disentangling content and style information and combining them within workflows of image reconstruction and transformation. Each image domain, denoted by $i\in\left\{ \mathcal{X},\mathcal{Y}\right\}$ is associated with a distinct set of modules. Each set consists of a content encoder ($E_{i}^{C}$), a style encoder ($E_{i}^{S}$), and a decoder ($D_i$). Fig. \ref{fig2} provides a visual overview of the network architecture for cross-domain image translation. Note that while the overall structure is capable of processing both complete images and image patches, the network parameters are trained using overlapped image patches. Data augmentation techniques applied to these patches, such as random rotation and flipping, enhance the model's performance. Therefore, to maintain clarity and consistency, the notations for image patches $X$ and $Y$ are used consistently throughout this section.

\subsubsection{Content Encoder}
Tasked with extracting style-invariant content features, the content encoder consists of five convolutional layers with a kernel size of 3 $\times$ 3, a stride of 1, and padding of 1, ensuring the output feature maps retain the same height and width as the input. Each layer is followed by an activation function. This design preserves spatial information crucial for accurate image reconstruction. The specific structure settings are shown in Table \ref{tab:structure_CE}. For content encoding process, we have:
\begin{equation}
\begin{split}
    &C_X=E_{\mathcal{X}}^{C}(X) 
    \\
    &C_Y=E_{\mathcal{Y}}^{C}(Y)
\end{split}
\end{equation}
where $C_X$ and $C_Y$ present the content codes corresponding to $X$ and $Y$.
\begin{table}[tbp]
    \centering
    \caption{Structure Details of Content Encoder\label{tab:structure_CE}}
    \renewcommand\arraystretch{1.2}
    \begin{tabular}{c|cccc}
    \hline
    \multirow{6}{*}{\begin{tabular}[t]{@{}c@{}}$E_\mathcal{X}^{C}$ \\ \vspace{5pt} $(E_\mathcal{Y}^{C})$\end{tabular}} & Layer Type & Filters & Kernel & Output \\ \cline{2-5} 
     & Input & - & - & $64 \times 64 \times C_1(C_2)$ \\ \cline{2-5}
     & Conv+Relu & 32 & $3 \times 3$ & $64 \times 64 \times 32$ \\
     & Conv+Relu & 64 & $3 \times 3$ & $64 \times 64 \times 64$ \\
     & Conv+Relu & 128 & $3 \times 3$ & $64 \times 64 \times 128$ \\
     & Conv+Relu & 128 & $3 \times 3$ & $64 \times 64 \times 128$ \\
     & Conv+Tanh & 128 & $3 \times 3$ & $64 \times 64 \times 128$ \\ \hline
    \end{tabular}
\end{table}

\subsubsection{Style Encoder}
Complementing the content encoder, the style encoder aims to capture the global stylistic attributes of the input image. Thus, we get style codes $S_X$ and $S_Y$ by:
\begin{equation}
    \begin{split}
        &S_X=E_{\mathcal{X}}^{S}(X)
        \\
        &S_Y=E_{\mathcal{Y}}^{S}(Y) 
    \end{split}
\end{equation}
This is accomplished using a sequence of four downsampling convolutional layers with a consistent kernel size of 3 $\times$ 3, a stride of 2, and padding of 1. These layers progressively reduce spatial resolution while expanding the receptive field. The sequence concludes with a global average pooling layer to derive a condensed style representation. Table \ref{tab:structure_SE} illustrates the detailed configuration of the style encoder.
\begin{table}[t!]
    \centering
    \caption{Structure detail of Style Encoder\label{tab:structure_SE}}
    \renewcommand\arraystretch{1.2}
    \begin{tabular}{c|cccc}
    \hline
    \multirow{7}{*}{\begin{tabular}[t]{@{}c@{}}$E_\mathcal{X}^{S}$ \\ \vspace{3pt} $(E_\mathcal{Y}^{S})$\end{tabular}} & Layer Type & Filters & Kernel & Output \\ \cline{2-5} 
     & Input & - & - & $64 \times 64 \times C_1(C_2)$ \\ \cline{2-5} 
     & Conv+Relu & 32 & $3 \times 3$ & $32 \times 32 \times 32$ \\
     & Conv+Relu & 64 & $3 \times 3$ & $16 \times 16 \times 64$ \\
     & Conv+Relu & 128 & $3 \times 3$ & $8 \times 8 \times 128$ \\
     & Conv+Relu & 256 & $3 \times 3$ & $4 \times 4 \times 256$ \\
     & AdaptiveAvgPool & - & - & $1 \times 1 \times 256$ \\ \hline
    \end{tabular}
\end{table}

\subsubsection{Decoder}
The decoder of CSTN demonstrates flexibility by performing both image reconstruction and image translation. When presented with content codes from the corresponding domain, it performs image reconstruction as:
\begin{equation}
\begin{split}
\label{eq:reconstruct}
&\tilde{X}=D_{\mathcal{X}}( C_X,S_X) =D_{\mathcal{X}}( E_{\mathcal{X}}^{C}( X ) ,E_{\mathcal{X}}^{S}( X )) 
\\
&\tilde{Y}=D_{\mathcal{Y}}(C_Y,S_Y) =D_{\mathcal{Y}}( E_{\mathcal{Y}}^{C}(Y) ,E_{\mathcal{Y}}^{S}( Y ) ) 
\end{split}
\end{equation}
Conversely, when provided with content codes from a different domain, it performs image transformation, effectively merging the input content with the target style. Formally, we have translated images as:
\begin{equation}
    \begin{split}
    \label{eq:translate}
        &\hat{X}=D_{\mathcal{X}}\left( C_Y,S_X \right) =D_{\mathcal{X}}\left( E_{\mathcal{Y}}^{C}\left( Y \right) ,E_{\mathcal{X}}^{S}\left( X \right) \right) 
        \\
        &\hat{Y}=D_{\mathcal{Y}}\left( C_X,S_Y \right) =D_{\mathcal{Y}}\left( E_{\mathcal{X}}^{C}\left( X \right) ,E_{\mathcal{Y}}^{S}\left( Y \right) \right) 
    \end{split}
\end{equation}

To inject style information into the decoding process, we utilize adaptive instance normalization (AdaIN)\cite{huang2017arbitrary} layers as formulated follows:
\begin{equation}
\label{eq_AdaIN}
AdaIN\left( z,\gamma ,\eta \right) =\gamma \left( \frac{z-\mu \left( z \right)}{\delta \left( z \right)} \right) +\eta  
\end{equation}
where $z$ represents the activations from the previous convolutional layer, $\mu$ and $\sigma$ denote the mean and standard deviation computed across each channel, respectively. The style code is processed by a multi-layer perceptron (MLP) with a structure of 1024-1024-8192 to generate the AdaIN parameters $\gamma$ and $\eta$ dynamically. 

\begin{figure}[tbp!]
	\centering
	\includegraphics[width=0.45\textwidth]{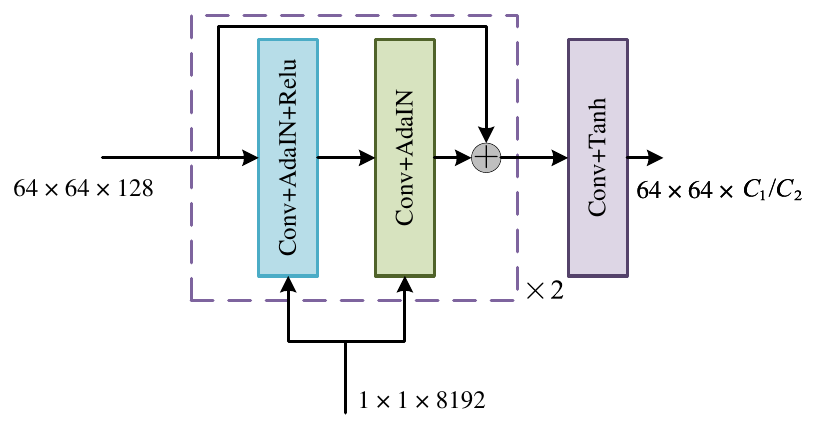}
	\caption{Structure of FFB.}
	\label{fig:feature_fusion_block}
\end{figure}

Apart from the MLP used for processing the style code, the decoder also includes a feature fusion block (FFB) designed to integrate both content and style information. The FFB consists of two residual blocks dedicated to feature mapping, complemented by a convolutional layer for dimension adjustment. Each residual block in this module contains two convolutional layers: the first is followed by an AdaIN layer and an activation function, while the second is followed solely by an activation function. All convolutional kernels within FFB are sized $3$ $\times$ $3$, and operate with padding and stride set to 1. Detailed insights into the structure of FFB are depicted in Fig. \ref{fig:feature_fusion_block}.

\subsection{Loss Functions}
\label{subsection:lossfunction}
Our model leverages a combination of reconstruction loss, translation loss, cycle consistency loss and content code alignment loss. This approach ensures the effective separation of content and style, precise image reconstruction and translation, and robust MCD.  
For all of them, we use the mean squared error (MSE) to calculate the loss between two matrices. For a matrix ($A$), the mean of all elements is denoted as $\mathbb{E}[A]$ and calculated as follows:
\begin{equation}
\mathbb{E}[A] = \frac{1}{mnl} \sum_{u=1}^{m} \sum_{v=1}^{n}\sum_{p=1}^{l} a_{uvp}
\end{equation}
The following subsections will delve into a detailed description of each loss function.

\subsubsection{Reconstruction Loss}
For images of a given modality, our model should be capable of decoding and reconstructing the images after decomposing them into content and style codes. To ensure this, the specific calculation of the loss is formulated as follows: 
\begin{equation}
\label{eq:reconloss}
\mathcal{L} _{recon}=\mathbb{E} \left[ \left( X-\hat{X} \right) ^2 \right] +\mathbb{E} \left[ \left( Y-\hat{Y} \right) ^2 \right] 
\end{equation}
where $X$ and $Y$ denote the original image patches,  $\hat{X}$ and $\hat{Y}$, represent the reconstruction results.

\subsubsection{Translation Loss}
In the cross-domain translation workflow, the proposed model aims to transform an input image into a translated image with the same modality of the target image by integrating the content code from the source domain and the style code from the target domain. Then, the translated images should be encoded to recover the content and style codes used in the translation process, which is formulated as follows:
\begin{equation}
    \begin{split}
    &\tilde{C}_X=E_{\mathcal{Y}}^{C}(\hat{Y})=E_{\mathcal{Y}}^{C}(D_{\mathcal{Y}}(C_X,S_Y))\cong C_X
    \\
    &\tilde{S}_Y=E_{\mathcal{Y}}^{S}(\hat{Y})=E_{\mathcal{X}}^{S}(D_{\mathcal{Y}}(C_X,S_Y))\cong S_Y
    \\
    &\tilde{C}_Y=E_{\mathcal{X}}^{C}(\hat{X})=E_{\mathcal{X}}^{C}(D_{\mathcal{X}}(C_Y,S_X))\cong C_Y
    \\
    &\tilde{S}_X=E_{\mathcal{X}}^{S}(\hat{X})=E_{\mathcal{X}}^{S}(D_{\mathcal{X}}(C_Y,S_X))\cong S_X
    \end{split}
\end{equation}
where $\tilde{C}_X$, $\tilde{S}_X$, $\tilde{C}_Y$ and $\tilde{S}_Y$ represents the reconstruction of content and style codes for $X$ and $Y$. Here, the symbol ``$\cong$" indicates that the two matrices are approximately equal, and this symbol will be used to represent the same meaning throughout this work. The final translation loss is the combination of content reconstruction and style reconstruction loss, and is formulated as: 
\begin{equation}
\begin{split}
\mathcal{L}_{trans}&=\mathbb{E} \left[ \left( C_X-\tilde{C}_X \right) ^2 \right]+\mathbb{E} \left[ \left( C_Y-\tilde{C}_Y \right) ^2 \right]   
\\
&+\mathbb{E} \left[ \left( S_X-\tilde{S}_X \right) ^2 \right] +\mathbb{E} \left[ \left( S_Y-\tilde{S}_Y \right) ^2 \right] 
\end{split}
\end{equation}

\subsubsection{Cycle Consistency Loss}
This cycle consistency loss ensures that translating an image to a different domain and then back to its original domain recovers the original image. It measures the discrepancy between the original image and its reconstruction after this two-step translation process. The loss is defined as:
\begin{equation}
\label{Cycloss_eq}
\begin{split}
\mathcal{L} _{cyc}=\mathbb{E} \left[ \left( X-\dot{X} \right) ^2 \right]+
+\mathbb{E} \left[ \left( Y-\dot{Y} \right) ^2 \right] 
\end{split}
\end{equation}
where $\dot{X}$ and $\dot{Y}$ represent the reconstructions of $X$ and $Y$ after image translation. To be more specific, the reconstruction processes after translation are defined as follows: 
\begin{equation}
    \begin{split}
        &\dot{X}=D_{\mathcal{X}}(\tilde{S}_X,\tilde{C}_X)\cong X
        \\
        &\dot{Y}=D_{\mathcal{Y}}(\tilde{S}_Y,\tilde{C}_Y)\cong Y
    \end{split}
\end{equation}

\subsubsection{Content Code Alignment Loss}
Under the assumption that content codes are invariant to style, content features of unchanged regions of the two domains should exhibit high similarity, while those of changed regions should present higher feature differences. Inspired by this, we introduce a mask-guided alignment strategy. A binary change mask is incorporated, denoted as $P_c$ where $P_c( i,j ) =1$ for changed pixels,  $P_c( i,j ) =0$ for unchanged ones. This mask mechanism guides the model to minimize the distance between content codes of unchanged regions while maximizing the distance between those of changed regions, thereby highlighting ground changes.
Furthermore, considering that the encoder-decoder architecture can reconstruct the content codes involved in the translation process, we impose an additional constraint. Specifically, we enforce cross-domain content consistency by minimizing the distance between $C_X$  and $\tilde{C}_Y$, $C_Y$  and $\tilde{C}_X$ during content code alignment, leveraging the property that $C_i$ and $\tilde{C}_i$ should be equivalent under the translation loss. 
For a given pixel$(i,j)$, $P_u(i,j)=1-P_c(i,j)$. This content code alignment loss further strengthens the model's ability to separate content from style, which is shown as follows: 
\begin{equation}
\label{eq:content_align}
\begin{split}
\mathcal{L} _{align}&=\mathbb{E} \left[ \left( C_X-\tilde{C}_Y \right) ^2\cdot P_u \right]+\mathbb{E} \left[ \left( \tilde{C}_X-C_Y \right) ^2\cdot P_u \right]
\\
&+\mathbb{E} \left[ \left( 1-\frac{\left( C_X-\tilde{C}_Y \right) ^2}{m} \right) \cdot P_c \right] 
\\
&+\mathbb{E} \left[ \left( 1-\frac{\left( \tilde{C}_X-C_Y \right) ^2}{m} \right) \cdot P_c \right] 
\end{split}
\end{equation}
where $m$ is a scaling parameter that adjusts the magnitude of the squared differences between matrices involved in the calculations. Since the final layer of the content encoder uses a hyperbolic tangent activation function, each component of the matrix $C_i$ approaches the range $(-1, 1)$. Consequently, the squared difference of each component between content codes falls within $(0, 4)$. We set $m$ to 4 to normalize these squared differences. This confines their values to the range of $(0, 1)$, which promotes numerical stability and supports effective learning.

The total loss function is listed as follows:
\begin{equation}
\label{eq:totalloss}
    \mathcal{L}_{total}=\mathcal{L}_{recon}+\mathcal{L}_{trans}+\mathcal{L}_{cyc}+\mathcal{L} _{align}
\end{equation}
Minimizing the loss function enables the joint training of all the network parameters $\bm{\theta}$. By assigning equal weights to all losses, the training process is simplified, reducing the need for hyperparameter tuning. This enhances the model's generalization across different scenarios. 

\subsection{Optimization}
An iterative refinement strategy is employed to optimize both the model parameters $\bm{\theta}$ and the change mask $P_c$ during training. At the beginning, $\bm{\theta}$ is initialized using PyTorch's default scheme, while each element of $P_c$ is randomly set to 0 or 1 according to a discrete uniform distribution. The optimization process alternates between two phases:

\begin{algorithm}[t!]
\caption{Procedure of CSTN}
\label{alg:algorithm1}
\begin{algorithmic}[1]
\renewcommand{\algorithmicrequire}{\textbf{Input:}}
\renewcommand{\algorithmicensure}{\textbf{Output:}}
\REQUIRE Pre-event Image $I_\mathcal{X}$, Post-event Image $I_\mathcal{Y}$, Number of Iterations $S$
\ENSURE Binary Change Map CM
\STATE Randomly initialize the network parameters $\bm{\theta}$ and change mask $P_c$.
\FOR{$s = 1$ to $S$}
    \FOR{$e=1$ to maximum epoch}
        \STATE Fix $P_c$, update network parameters $\bm{\theta}$ through minimizing Eq. (\ref{eq:totalloss}).
    \ENDFOR
    \STATE Fix $\bm{\theta}$, get whole image content codes and compute the difference image.
    \STATE Obtain CM by Eq. (\ref{threshold_eq}), update the change mask $P_c$.
\ENDFOR
\STATE Filter the final DI, and obtain the CM by thresholding.
\RETURN CM
\end{algorithmic}
\end{algorithm}

\subsubsection{Parameter Update}
In this phase, the change mask $P_c$ is held constant, and the model parameters $\bm{\theta}$ is updated via backpropagation by minimizing the overall loss function. Notably, these loss components do not require additional weight hyperparameters for balancing.

\subsubsection{Mask Update}
\label{mask_update}
After updating the model parameters, $\bm{\theta}$ is fixed, and the change mask $P_c$ is further refined. By utilizing the entire image as input of the cross-domain image translation and cycle-reconstruction workflow, we obtain content codes ($C_\mathcal{X}$ and $C_\mathcal{Y}$) for both images, along with reconstructed content codes ($\tilde{C}_\mathcal{X}$ and $\tilde{C}_\mathcal{Y}$) derived from the translated images. Building upon the alignment employed in the content code alignment loss, a difference image can be generated by measuring the distance between each position of $C_\mathcal{X}$ and $\tilde{C}_\mathcal{Y}$, $C_\mathcal{Y}$ and $\tilde{C}_\mathcal{X}$. To efficiently fuse these complementary information and simplify computation, we concatenate the content codes along the channel dimension, resulting in $f_\mathcal{X}=concat(C_\mathcal{X}, \tilde{C}_\mathcal{X})$ and $f_\mathcal{Y}=concat(\tilde{C}_\mathcal{Y},C_\mathcal{Y})$. Treating these concatenated matrices as image representations, we compute the Euclidean distance between corresponding pixels, yielding the difference image (DI) as follows:
\begin{equation}
    \text{DI}(i,j)=\left\| f_\mathcal{X}\left( i,j \right) -f_\mathcal{Y}\left( i,j \right) \right\| _2
\end{equation}
This approach facilitates a comprehensive comparison between the original and reconstructed content codes, which results in the DI. We then employ the Otsu threshold method \cite{otsu1979threshold} to segment DI and get the change map (CM), effectively distinguishing changed pixels from unchanged ones. Pixels with values exceeding the calculated threshold $T$, are classified as changed (assigned a value of 1), while those below the threshold are designated as unchanged (assigned a value of 0), as formalized below:
\begin{equation}
\label{threshold_eq}
\text{CM}(i,j) = 
\begin{cases} 
1, & \text{if } \text{DI}(i,j) > T \\ 
0, & \text{if } \text{DI}(i,j) \leq T 
\end{cases}
\end{equation}

This iterative refinement procedure, alternating between model parameters $\bm{\theta}$ and change mask $P_c$ updates, allows the model to progressively improve its ability to separate content and style, consequently enhancing image translation and change detection performance. Algorithm \ref{alg:algorithm1} details the iterative updating process.

\subsection{Change Detection Pipeline}
Once the model is trained, a change detection pipeline is performed to generate accurate detection results through a straightforward feature comparison manner. First, following the previously described workflow, we obtain a DI highlighting potential changes. The DI, however, may contain noise and spurious detection. To mitigate these artifacts, we adopt Gaussian filtering \cite{zhang2012efficient}, which leverages spatial context to regularize the difference map and reduce erroneous classifications. Finally, we apply a threshold as described in Section \ref{mask_update} to the filtered DI to generate the final binary CM.

\begin{table*}[ht!]
    \centering
    \caption{Descriptions of The Four Multimodal Datasets\label{tab:table1}}
    \renewcommand\arraystretch{1.5}
    \begin{tabular}{cccccc}
    \hline
        \textbf{Dataset} & \textbf{Imaging Sensor } & \textbf{Image Size } & \textbf{Date } & \textbf{Location } & \textbf{Event(Spatial Resolution)} \\ \hline
        Sardinia  & Landsat-5/Google Earth  & 300$\times$412$\times$1(3)  & Sep. 1995/Jul. 1996 & Sardinia, Italy  & Lake Expansion(30m) \\ 
        Texas  & Landsat-5/EO-1 ALI  & 1534$\times$808$\times$7(10)  & Stp. 2011/Oct. 2011 & Texas, USA  & Forest Fire(30m) \\ 
        Shuguang  & Radarsat-2/Google Earth  & 593$\times$921$\times$1(3)  & Jun. 2008/Sep. 2012 & Shuguang Village, China  & Building Construction(8m) \\ 
        California  & Landsat-8/Sentinel-1A  & 3500$\times$2000$\times$11(3) & Jan. 2017/Feb. 2017 & Sutter County, USA  & Flooding($\approx$15m) \\ \hline
    \end{tabular}
\end{table*}

\begin{figure}[t!]
\includegraphics[width=0.45\textwidth]{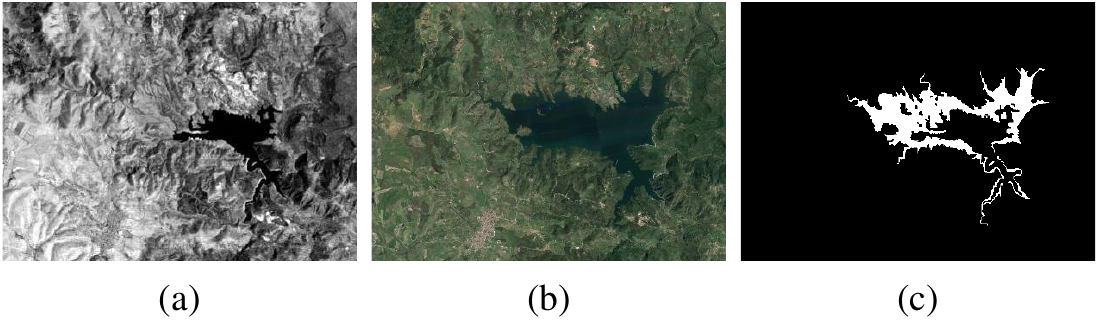}
\caption{Sardinia dataset. (a) Near-infrared image acquired in 1995. (b) Optical image acquired in 1996. (c) Ground truth.}
\label{fig:dataset_sardinia}
\end{figure}

\begin{figure}[t!]
\includegraphics[width=0.45\textwidth]{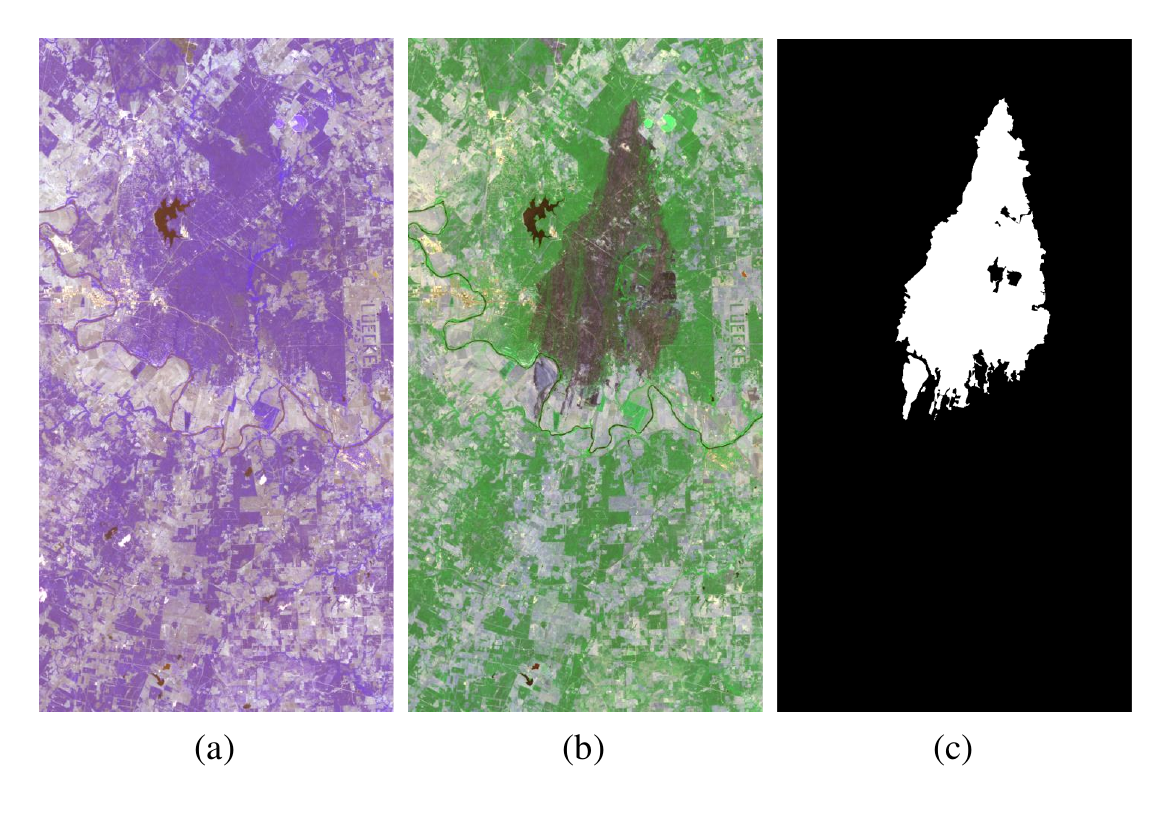}
\caption{Texas dataset. (a) Optical image acquired in 2011. (b) Optical image acquired in 2011. (c) Ground truth. False color composites are shown for both the images.}
\label{fig:dataset_Texas}
\end{figure}
\section{Experiments Results and Analysis}
\label{Section:experiments}
\subsection{Datasets}
\begin{figure}[t!]
\includegraphics[width=0.45\textwidth]{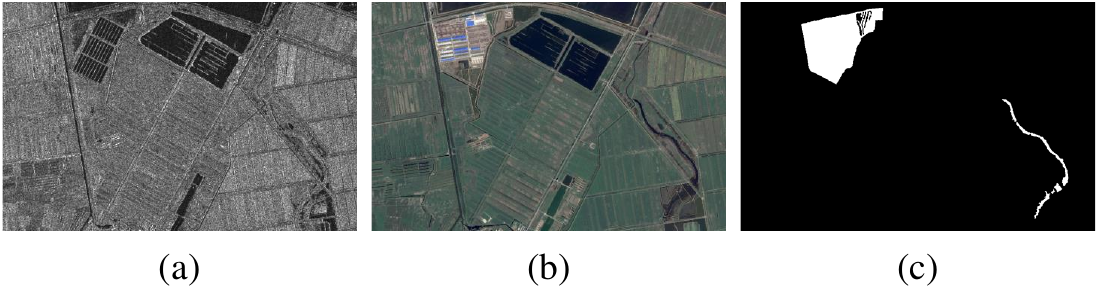}
\caption{Shuguang dataset. (a) SAR image acquired in 2008. (b) Optical image acquired in 2012. (c) Ground truth. }
\label{fig:dataset_shuguang}
\end{figure}

\begin{figure}[t!]
\includegraphics[width=0.45\textwidth]{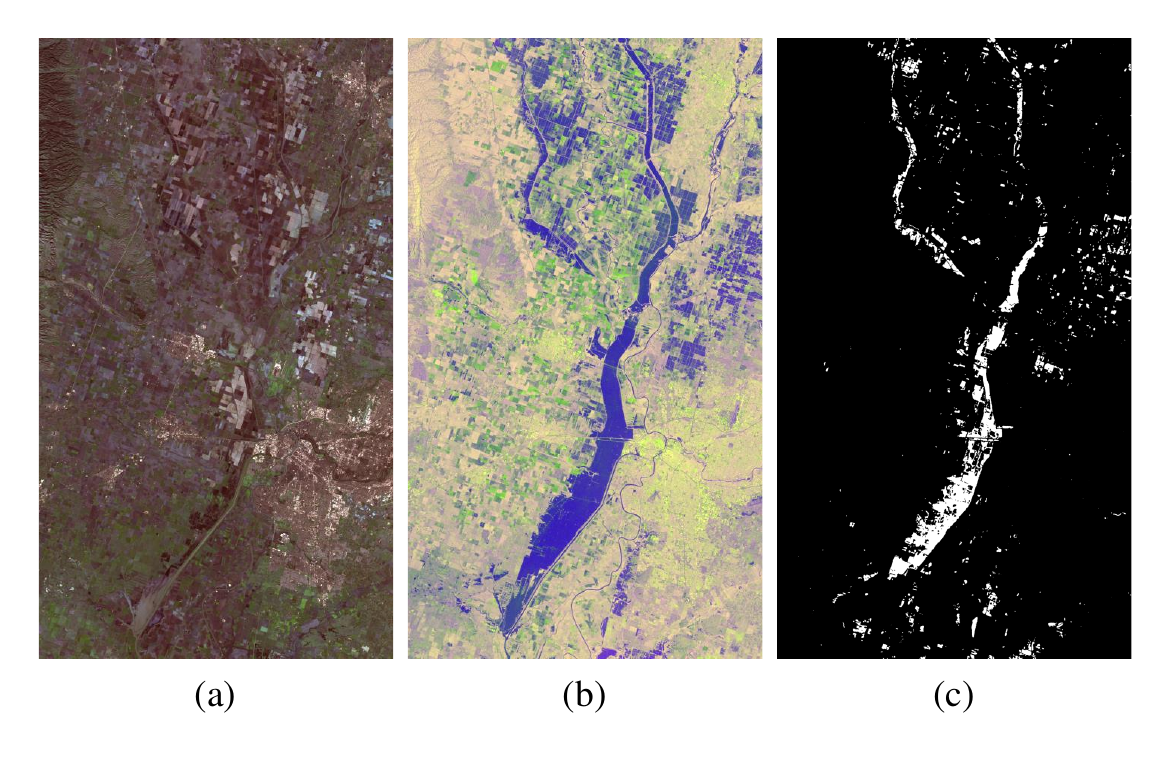}
\caption{California dataset. (a) Optical image acquired in 2017. (b) SAR image acquired in 2017. (c) Ground truth. False color composites are shown for both the images.}
\label{fig:dataset_California}
\end{figure}

To rigorously evaluate the effectiveness and generalization capabilities of the proposed framework, we conducted experiments on four multimodal remote sensing datasets which are shown in Figs. \ref{fig:dataset_sardinia} - \ref{fig:dataset_California}. Each dataset was carefully selected to encompass a wide range of data characteristics and change events, ensuring a comprehensive assessment of the model's performance across diverse scenarios. These datasets vary significantly in terms of image types, including SAR, visible RGB (red, green, blue), near-infrared, and multispectral images. Moreover, they exhibit diverse image sizes, ranging from 300 to 3500 pixels in length or width, and spatial resolutions, spanning from 8 m to 30 m. Furthermore, they cover a variety of change events, including but not limited to flooding and fire, further enhancing their diversity. Table \ref{tab:table1} provides a detailed summary of each dataset, highlighting their specific attributes. These datasets are widely recognized benchmarks for MCD tasks. The ground truth annotations are manually created, integrating expert knowledge with background information.

\subsection{Experiments Settings}
\subsubsection{Implementation Details}
During the phase of model training, the input image was first cropped into patches of size 64 $\times$ 64 with a stride of 56 pixels. These patches were then used to train the model using the Adam optimizer\cite{2014Adam} with an initial learning rate of 0.0001 and momentum parameters $\beta = \{0.5, 0.9\}$. The mini-batch size was set as 32. For each iteration, the model was trained for 10 epochs, ensuring that the entire training dataset was processed in each epoch. Following that, $P_c$ was updated using the entire image as input. This process of alternating between network parameters $\bm{\theta}$ and change mask $P_c$ updates was repeated for a total of 2 iterations. Additionally, within the implementation details, the original California dataset, which initially comprised images with dimensions of 3500 $\times$ 2000 pixels, was resampled to 875 $\times$ 500 pixels to align with common practices adopted by many other studies.

Based on the aforementioned categorization of MCD methods, we selected representative unsupervised approaches from three categories, excluding classification-based method, which primarily consists of supervised methods. Specifically, we chose INLPG\footnote{\url{https://github.com/yulisun/INLPG}}\cite{sun2021structure} and IRG-McS\footnote{\url{https://github.com/yulisun/IRG-McS}}\cite{sun2021iterative} from handcrafted feature-based methods, SCCN\cite{liu2016deep} and PMBCN\footnote{\url{https://github.com/liusiqinqinqin/probabilistic-model-based-changedetection}}\cite{liu2021probabilistic} from deep latent feature-based methods, and CAN\cite{niu2018conditional}, X-Net\footnote{\url{https: //github.com/llu025/Heterogeneous_CD}\label{fn_4_x_ACE_CAAE}}\cite{luppino2021deep}, ACE-Net\footref{fn_4_x_ACE_CAAE}\cite{luppino2021deep}, CAAE\footref{fn_4_x_ACE_CAAE}\cite{luppino2022code} and SRF\footnote{\url{https://github.com/yulisun/SRF}}\cite{sun2023structural} from image translation-based methods. Most of these methods are publicly available, and we utilized their official implementations. For SCCN and CAN, we implemented them in PyTorch based on the details provided in the original papers. All experiments were conducted on a workstation equipped with an NVIDIA GeForce RTX 4090 D GPU (24 GB VRAM) and an AMD Ryzen 9 7950X 16-Core CPU, featuring a maximum clock speed of 5.88 GHz.

\begin{figure*}[ht]
	\centering
\includegraphics[width=0.95\textwidth]{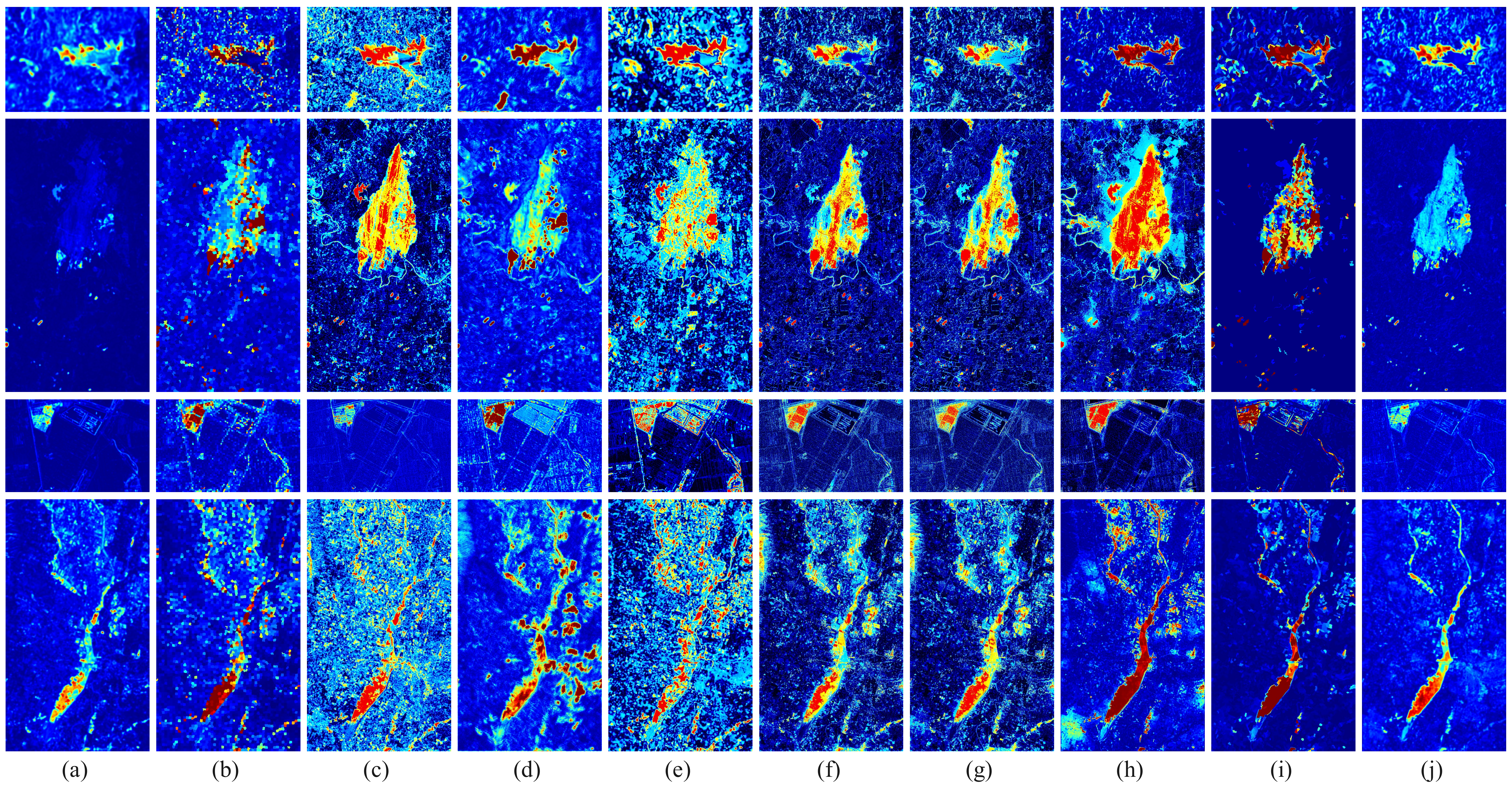}
	\caption{DIs generated by different change detection methods on the four datasets. (a) INLPG; (b) IRG-McS; (c) SCCN; (d) PMBCN; (e) CAN; (f) X-Net; (g) ACE-Net; (h) CAAE; (i) SRF; (j) CSTN. From top to bottom, displayed are DIs in the form of heat map across four datasets: Sardinia, Texas, Shuguang and California. Brighter red areas indicate a higher likelihood of change, while darker blue areas suggest a higher probability of remaining unchanged.}
	\label{fig:DI}
\end{figure*}

\subsubsection{Evaluation Metrics}
We evaluated the performance of our method and competing methods using three types of metrics. First, to assess the quality of DI, we employed Receiver Operating Characteristic (ROC) curves and Precision-Recall (PR) curves. A higher quality DI is indicated by an ROC curve closer to the upper left corner, a PR curve closer to the upper right corner, and consequently, higher values for the Area Under the Curve (AUC) and Average Precision (AP).
Second, to evaluate CM, we visually represent True Positive (TP), False Positive (FP), True Negative (TN), and False Negative (FN) using distinct colors. Additionally, we utilize six widely adopted metrics \cite{luppino2022code} to quantitatively analyze the results: FN, FP, Overall Error (OE), Overall Accuracy (OA), F1 Score (F1), and Kappa Coefficient (KC). Lower values for FP, FN, and OE, along with higher values for OA, F1, and KC, indicate better change detection performance. To further ensure a comprehensive evaluation of our model, we employed quantitative metrics to assess the quality of image translation. Specifically, we utilized the Fréchet Inception Distance (FID) \cite{heusel2017gans} and Kernel Inception Distance (KID) \cite{Bińkowski2018Demystifying} to quantify the discrepancy between real and translated images. Both FID and KID leverage features extracted from the penultimate layer of a pre-trained Inception V3 network \cite{szegedy2016rethinking}. The lower values of two indicators mean better translation results. Their calculations are shown in Eq. (\ref{eq:FID}) and Eq. (\ref{eq:KID}).
\begin{equation}
\label{eq:FID}
    \text{FID} = \|\mu_r - \mu_t\|^2_2 + \text{Tr}(\Sigma_r + \Sigma_t - 2(\Sigma_r \Sigma_t)^{1/2})
\end{equation}
where $ \mu_r $ and $ \mu_t $ are the mean vectors of the real and translated images, respectively, and $ \Sigma_r $ and $ \Sigma_t $ are their covariance matrices. $ \text{Tr} $ represents the trace of the matrix, and $ \|\cdot\|^2_2 $ denotes the squared Euclidean distance between the means.
\begin{equation}
\label{eq:KID}
\begin{split}
        \text{KID} = & \mathbb{E}_{r \sim P_{\text{real}}} [k(r, r')] + \mathbb{E}_{t \sim P_{\text{trans}}} [k(t, t')] 
        \\
        & - 2\mathbb{E}_{r \sim P_{\text{real}}, t \sim P_{\text{trans}}} [k(r, t)]
\end{split}
\end{equation}
In this context, $ r $ and $r'$ denote samples from the set of real images, while $t$ and $ t'$ represent samples from the set of generated images. In our specific computation, $r'=r$ and $t'=t$. The polynomial kernel function $ k $ is specified as follows:
\begin{equation}
     k(r, t) = \left(\frac{1}{d} r^T t + 1\right)^3 
\end{equation}
where $ d $ represents the dimension of the image $ r $.

\subsection{Analysis for DI}
\begin{table*}[htbp!]
    \centering
    \caption{The Values of AUC and AP of DIs Obtained by Different Methods on the Four Datasets. The Best Results are in Bold, and the
    Second-best Results are Underlined.\label{tab:table2}}
    \renewcommand\arraystretch{1.2}
\begin{tabular}{c|c|cccccccccc}
\hline
Dataset & Indicator & INLPG & IRG-McS & SCCN & PMBCN & CAN & X-Net & ACE-Net & CAAE & SRF & CSTN \\ \hline
\multirow{2}{*}{Sardinia} & AUC & {\ul 0.9503} & 0.8985 & 0.9368 & 0.9294 & 0.9473 & 0.9130 & 0.9070 & \textbf{0.9640} & 0.9443 & 0.9469 \\
 & AP & 0.7206 & 0.6428 & 0.7740 & 0.6385 & {\ul 0.8181} & 0.6313 & 0.5837 & 0.8146 & 0.7565 & \textbf{0.8198} \\ \hline
\multirow{2}{*}{Texas} & AUC & 0.9755 & 0.9567 & 0.9896 & 0.9476 & 0.9510 & 0.9744 & 0.9805 & {\ul 0.9906} & 0.9642 & \textbf{0.9929} \\
 & AP & 0.7206 & 0.6579 & 0.8590 & 0.6629 & 0.6285 & 0.7864 & 0.8334 & {\ul 0.8692} & 0.8196 & \textbf{0.8974} \\ \hline
\multirow{2}{*}{Shuguang} & AUC & {\ul \textbf{0.9843}} & 0.9802 & 0.9436 & 0.9458 & 0.9536 & 0.9737 & 0.9650 & {\ul \textbf{0.9861}} & 0.9624 & 0.9798 \\
 & AP & 0.8111 & 0.7849 & 0.6867 & 0.7272 & 0.4566 & 0.7697 & 0.7069 & {\ul 0.8190} & {\ul 0.7753} & \textbf{0.8278} \\ \hline
\multirow{2}{*}{California} & AUC & {\ul \textbf{0.9386}} & 0.9265 & 0.8939 & 0.8638 & 0.8371 & 0.9159 & 0.9097 & {\ul \textbf{0.9282}} & 0.9271 & 0.9145 \\
 & AP & 0.4490 & 0.4419 & 0.4612 & 0.2061 & 0.2406 & 0.3325 & 0.3576 & {\ul 0.5049} & {\ul 0.5369} & \textbf{0.5394} \\ \hline
\end{tabular}
\end{table*}
\begin{figure*}[htbp]
    \centering
    \includegraphics[width=1\textwidth]{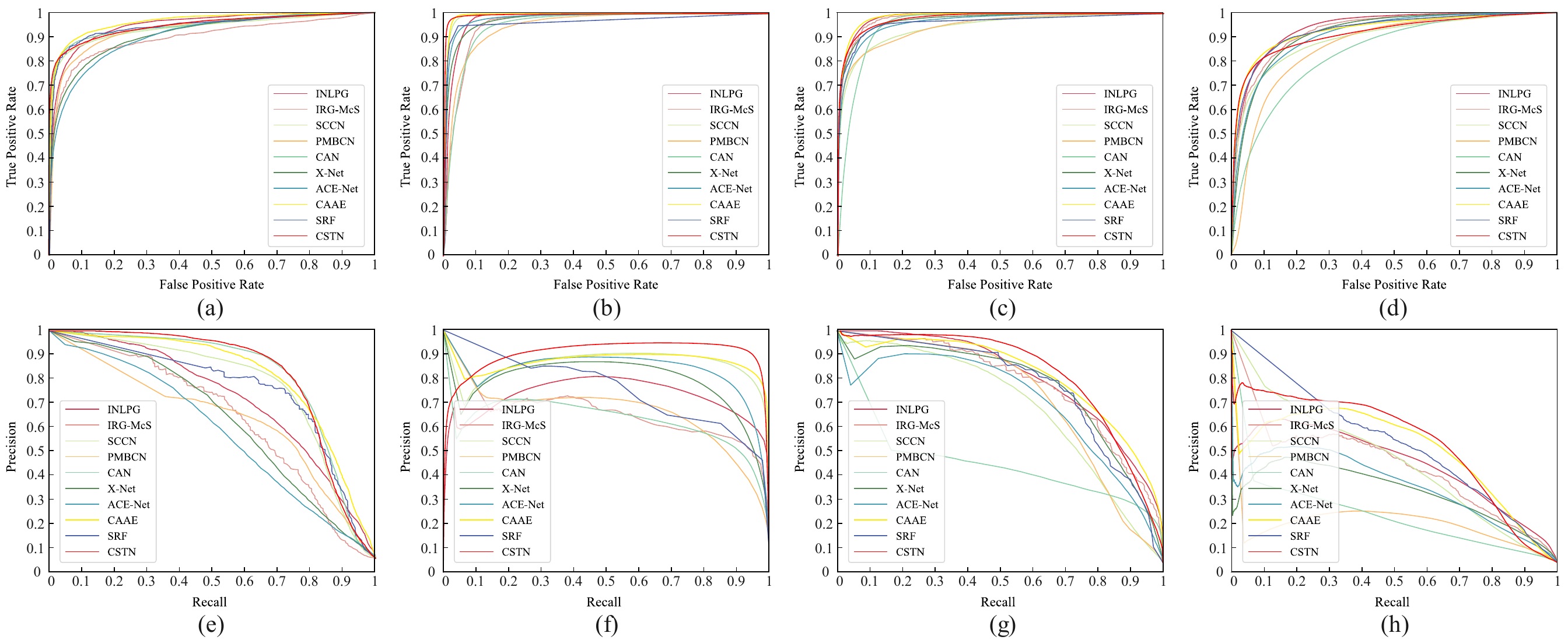}
    \caption{ROC and PR curves of DIs generated by different methods. (a)-(d) are the ROC curves, and (e)-(h) are the PR curves. From left to right, the results correspond to the Sardinia, Texas, Shuguang and California datasets, respectively.}
    \label{fig:ROC_PRS}
\end{figure*}
Since all the selected MCD methods follow the principle of generating a DI followed by change analysis, an effective DI should clearly differentiate between areas that have undergone change and those that remain unchanged. This section evaluates the quality of the proposed method and comparative algorithms in generating DI, employing both qualitative and quantitative assessments.

To enhance clarity and intuitively demonstrate its quality, we present DIs as heat maps in Fig. \ref{fig:DI}. Overall, the performance of different models varies across datasets, due to the different modalities each dataset contains. CSTN, though not showing the highest contrast between changed and unchanged areas, consistently delivers commendable results across all four datasets, distinguishing itself with balanced performance. In all DIs generated by CSTN, the changed and unchanged areas are clearly distinguishable as shown in Fig. \ref{fig:DI}(j). However, other models exhibit significant instability in performance across different datasets. Especially in complex scenes, such as the California dataset (the last row of Fig. \ref{fig:DI}), most methods show noise in the unchanged areas.

Then, by applying a simple threshold segmentation algorithm (i.e., the Otsu) on the above DIs, the quality of DIs can be analyzed by the ROC and PR curves in Fig. \ref{fig:ROC_PRS}. The quantitative evaluation results of DIs are shown in Table \ref{tab:table2}, CSTN demonstrated superior performance across multiple datasets, achieving the highest AP scores on the Sardinia (0.8198), Texas (0.8974), and Shuguang (0.8278) datasets, and the highest AUC on the Texas (0.9929) dataset. Similar to the visual results, the DIs of all models on the California dataset are limited by complex scenes, performing poorly; however, CSTN still achieved the highest AP score among all algorithms. Additionally, CSTN consistently achieved competitive AUC scores, underscoring its effectiveness and robustness in MCD tasks compared to other methods. The ROC and PR curves in Fig. \ref{fig:ROC_PRS} provide a visual confirmation of this performance difference.
\begin{figure}[t!]
	\centering
	\includegraphics[width=0.45\textwidth]{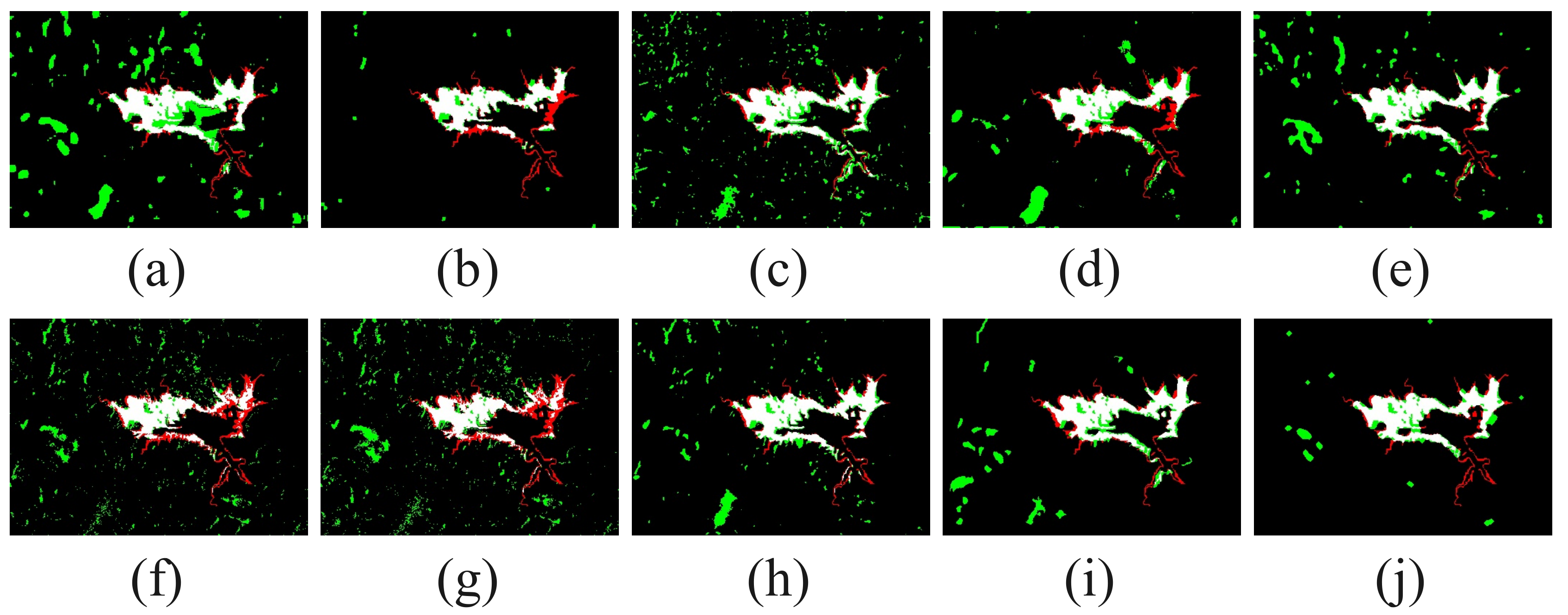}
	\caption{CMs of different methods on the Sardinia dataset. (a) INLPG; (b) IRG-McS; (c) SCCN; (d) PMBCN; (e) CAN; (f) X-Net; (g) ACE-Net; (h) CAAE; (i) SRF; (j) CSTN. (TP: white; TN: black; FP: green; FN: red).}
	\label{fig:CM_Sardinia}
\end{figure}
\begin{figure}[t!]
	\centering
	\includegraphics[width=0.45\textwidth]{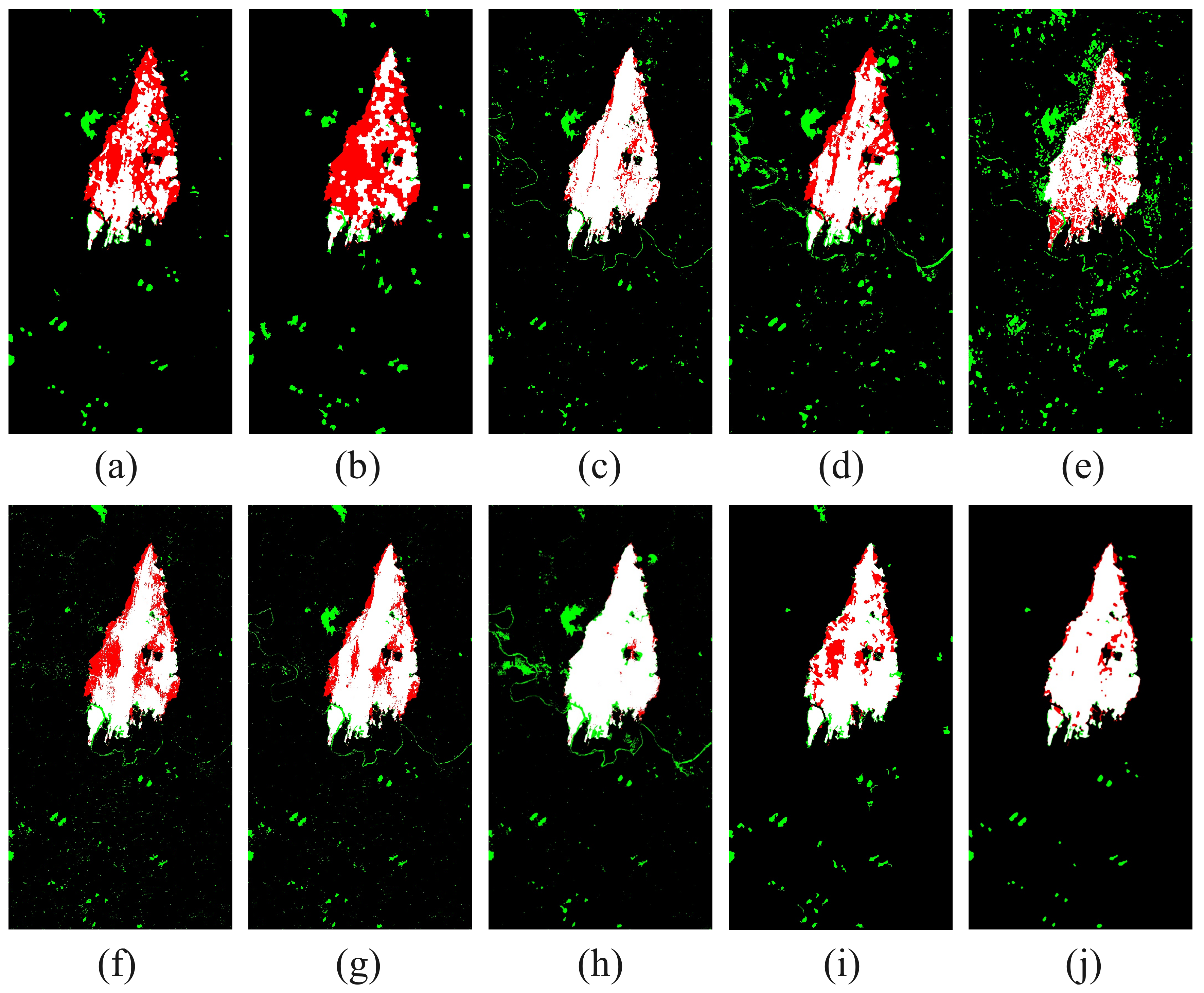}
	\caption{CMs of different methods on the Texas dataset. (a) INLPG; (b) IRG-McS; (c) SCCN; (d) PMBCN; (e) CAN; (f) X-Net; (g) ACE-Net; (h) CAAE; (i) SRF; (j) CSTN. (TP: white; TN: black; FP: green; FN: red).}
	\label{fig:CM_Texas}
\end{figure}
\subsection{Analysis of CM}
Our proposed CSTN outperforms existing methods on four multimodal datasets, achieving significant improvements based on OA, F1, and KC metrics. Qualitative visualizations of change maps further support this, demonstrating CSTN's ability to capture complex patterns and subtle data variations. This subsection focuses on a dataset-by-dataset analysis of CMs generated by all methods.

\subsubsection{Results on the Sardinia Dataset}
Fig. \ref{fig:CM_Sardinia} and Table \ref{tab:CM_Sarfinia} display the evaluation results of CMs for various models on the Sardinia dataset. Although the dataset is small, it contains numerous intricate details that challenge the models' ability to detect changes accurately. Several methods exhibit notable weaknesses. For example, INLPG shows a great amount of FP (with large green areas in Fig. \ref{fig:CM_Sardinia} (a)), coupled with relatively low F1 and KC, indicating limited accuracy and reliability. Similarly, SCCN, despite minimizing FN, exhibits a high OE and moderate F1 and KC, suggesting challenges in identifying true change pixels accurately. PMBCN and CAN achieve lower OE, yet their performance in terms of F1 and KC is inferior, revealing limitations in effectively discriminating between changed and unchanged pixels. In contrast, our proposed CSTN model outperforms other methods in OA, F1, and KC, demonstrating superior performance. This robust performance, validated both quantitatively and visually, underscores CSTN's advanced capability in MCD tasks.

\subsubsection{Results on the Texas Dataset}
The evaluation results of the Texas dataset, as shown in Fig. \ref{fig:CM_Texas} and Table \ref{tab:CM_Texas}, reveal the following insights. Handcrafted feature-based methods like INLPG and IRG-McS exhibit limited effectiveness, with notable red FN regions (see Fig. \ref{fig:CM_Texas}(a) and (b)). Deep latent feature-based methods such as SCCN present competitive results but still fall short of CSTN's performance across all metrics. Although CAAE demonstrates superiority over other image translation-based methods, it tends to misclassify some unchanged regions as changed, resulting in a higher incidence of FP despite a lower count of FN. Consequently, CAAE's overall F1 Score is approximately 5.1\% lower, and its KC is about 6.0\% lower than CSTN's. Among all methods, CSTN excels with the lowest OE and the highest OA, F1, and KC, which demonstrates its precision and robustness.

\begin{table}[t!]
    \centering
    \caption{Performance Comparison Between Different Change Detection Methods on the Sardinia Dataset. The Best Results are in Bold, and the Second-best are Underlined.\label{tab:CM_Sarfinia}}
\renewcommand\arraystretch{1.2}
\begin{tabular}{c|cccccc}
\hline
Method & FP & FN & OE & OA & F1 & KC \\ \hline
INLPG & 7772 & 1385 & 9157 & 0.9259 & 0.5768 & 0.5401 \\
IRG-McS & \textbf{1323} & 2214 & \textbf{3537} & \textbf{0.9714} & {\ul 0.7537} & {\ul 0.7386} \\
SCCN & 5862 & {\ul 1312} & 7174 & 0.9419 & 0.6377 & 0.6079 \\
PMBCN & 4069 & 2178 & 6247 & 0.9495 & 0.6356 & 0.6088 \\
CAN & 5043 & \textbf{1218} & 6261 & 0.9493 & 0.6718 & 0.6456 \\
X-Net & 3800 & 2895 & 6695 & 0.9458 & 0.5856 & 0.5567 \\
ACE-Net & 4589 & 3063 & 7652 & 0.9381 & 0.5439 & 0.5110 \\
CAAE & 3781 & 1348 & 5129 & 0.9585 & 0.7100 & 0.6881 \\
SRF & 3687 & 1471 & 5158 & 0.9583 & 0.7047 & 0.6827 \\
CSTN & {\ul 2192} & 1446 & {\ul 3638} & {\ul 0.9706} & \textbf{0.7726} & \textbf{0.7569} \\ \hline
\end{tabular}
\end{table}

\begin{table}[t!]
    \centering
    \caption{Performance Comparison Between Different Change Detection Methods on the Texas Dataset. The Best Results are in Bold, and the Second-best are Underlined.\label{tab:CM_Texas}}
\renewcommand\arraystretch{1.2}
\begin{tabular}{c|cccccc}
\hline
Method & FP & FN & OE & OA & F1 & KC \\ \hline
INLPG & 16467 & 63364 & 79831 & 0.9356 & 0.6318 & 0.5984 \\
IRG-McS & 21697 & 76747 & 98444 & 0.9206 & 0.5283 & 0.4882 \\
SCCN & 19928 & {\ul 14909} & {\ul 34837} & {\ul 0.9719} & 0.8704 & 0.8546 \\
PMBCN & 48178 & 38909 & 87087 & 0.9297 & 0.6810 & 0.6414 \\
CAN & 58121 & 40970 & 99091 & 0.9199 & 0.6472 & 0.6022 \\
X-Net & 19088 & 39271 & 58359 & 0.9529 & 0.7604 & 0.7345 \\
ACE-Net & 20585 & 26174 & 46759 & 0.9623 & 0.8189 & 0.7978 \\
CAAE & 32553 & \textbf{3790} & 36343 & 0.9707 & {\ul 0.8757} & {\ul 0.8593} \\
SRF & {\ul 14955} & 25472 & 40427 & 0.9674 & 0.8403 & 0.8222 \\
CSTN & \textbf{7970} & 12044 & \textbf{20014} & \textbf{0.9839} & \textbf{0.9229} & \textbf{0.9139} \\ \hline
\end{tabular}
\end{table}

\subsubsection{Results on the Shuguang Dataset}
\begin{figure}[t]
	\centering
	\includegraphics[width=0.45\textwidth]{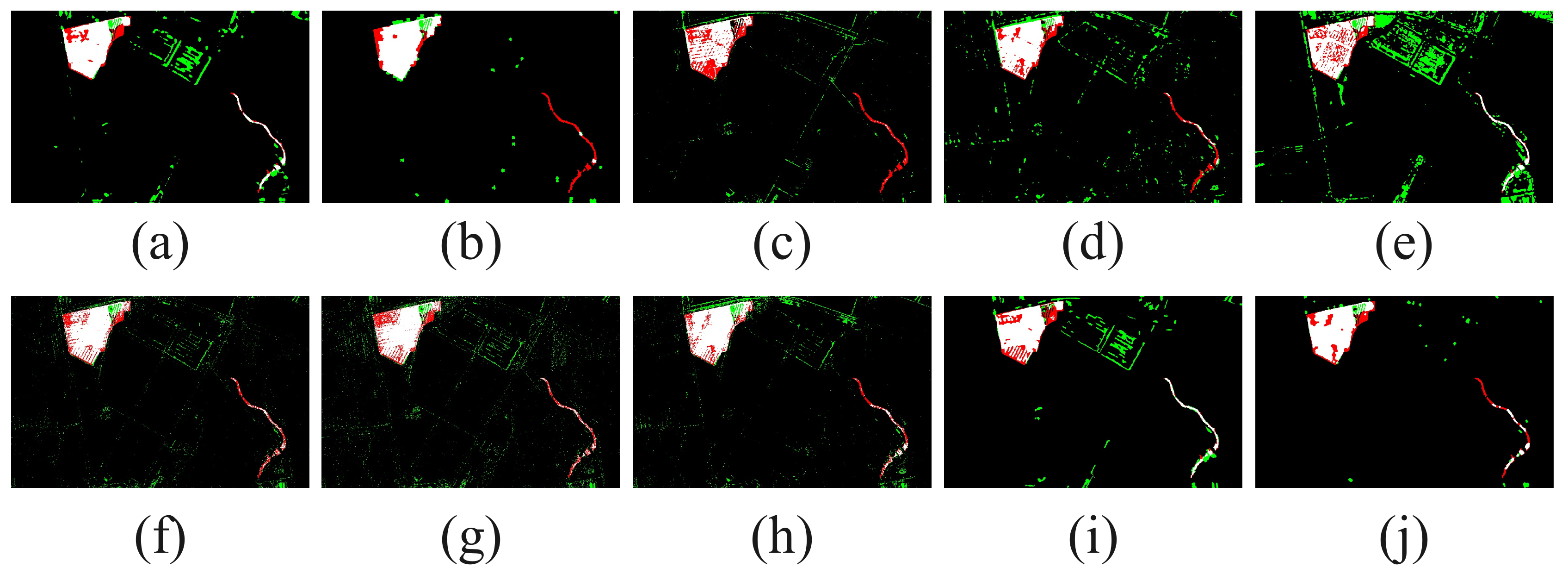}
	\caption{CMs of different methods on the Shuguang dataset. (a) INLPG; (b) IRG-McS; (c) SCCN; (d) PMBCN; (e) CAN; (f) X-Net; (g) ACE-Net; (h) CAAE; (i) SRF;  (j) CSTN. (TP: white; TN: black; FP: green; FN: red).}
	\label{fig:CM_Shuguang}
\end{figure}

The evaluation of various change detection methods on the Shuguang dataset (as shown in  Fig. \ref{fig:CM_Shuguang} and Table \ref{tab:CM_Shuguang}) underscores the exceptional performance of CSTN. While the handcrafted feature-based method IRG-McS achieves commendable results, including the second-lowest OE and high metrics, it still falls short compared to CSTN. The two deep latent feature-based methods we selected do not perform as well as other approaches, likely due to their limited feature extraction capabilities, which result in feature spaces less suited for change detection. Among the image translation-based methods, CAAE and X-Net demonstrate some success, but they still lag behind CSTN. Their change detection results remain suboptimal, possibly due to the influence of image translation quality. Additionally, as illustrated in Fig. \ref{fig:CM_Shuguang}, most models are influenced by the farmland adjacent to the change areas, as indicated by the prominent green contours. CSTN consistently delivers superior accuracy and reliability, as evidenced by both visual maps and performance metrics, firmly establishing it as the leading method in all evaluated aspects.

\subsubsection{Results on the California Dataset}
\begin{figure}[t!]
	\centering
	\includegraphics[width=0.45\textwidth]{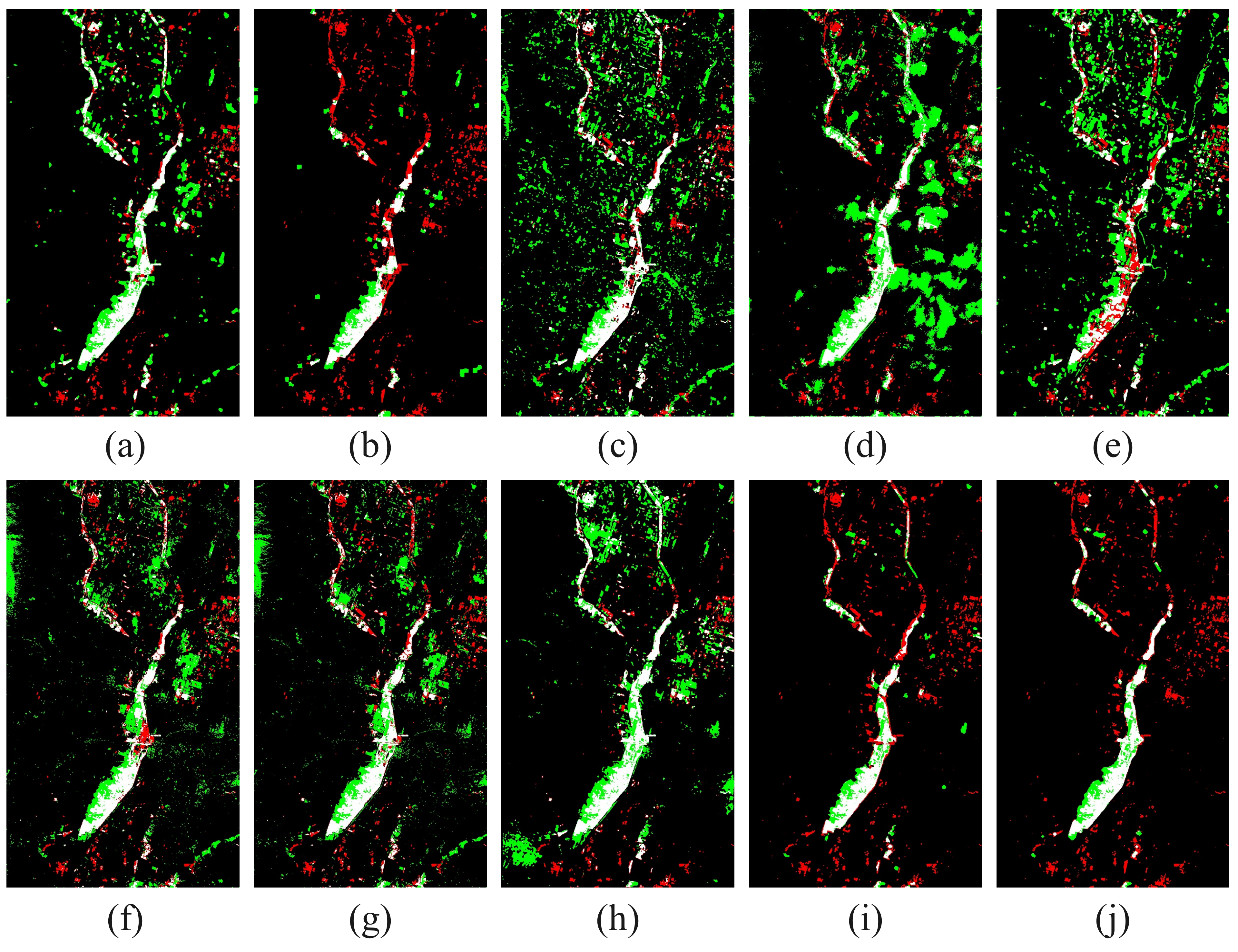}
	\caption{CMs of different methods on the California dataset. (a) INLPG; (b) IRG-McS; (c) SCCN; (d) PMBCN; (e) CAN; (f) X-Net; (g) ACE-Net; (h) CAAE; (i) SRF;  (j) CSTN. (TP: white; TN: black; FP: green; FN: red).}
	\label{fig:CM_California}
\end{figure}

The performance evaluation of various change detection methods on the California dataset is illustrated in  Fig. \ref{fig:CM_California} and Table \ref{tab:CM_California}. The complexity of the California dataset, which includes a greater variety of ground objects such as mountains, vegetation, and rivers, poses significant challenges compared to the previous three datasets. This increased complexity results in poorer performance across all evaluated models, as evidenced by notably lower KC values. Among the models, IRG-McS and SRF demonstrate competitive performance, with SRF achieving the second-best OE and OA. However, CSTN consistently outperforms both in critical metrics, including the F1 score and KC. This robust performance is also visually apparent, with CSTN maintaining a stable and balanced performance compared to the fluctuating FP (marked with green color) observed in other methods.


\begin{table}[t!]
    \centering
    \caption{Performance Comparison Between Different Change Detection Methods on the Shuguang Dataset. The Best Results are in Bold, and the Second-best are Underlined.\label{tab:CM_Shuguang}}
\renewcommand\arraystretch{1.2}
\begin{tabular}{c|cccccc}
\hline
Method & FP & FN & OE & OA & F1 & KC \\ \hline
INLPG & 12438 & \textbf{4666} & 17104 & 0.9687 & 0.7050 & 0.6887 \\
IRG-McS & 3889 & 5667 & {\ul 9556} & {\ul 0.9825} & {\ul 0.8026} & {\ul 0.7935} \\
SCCN & {\ul 5576} & 10408 & 15984 & 0.9707 & 0.6477 & 0.6326 \\
PMBCN & 12971 & 6909 & 19880 & 0.9636 & 0.6466 & 0.6277 \\
CAN & 33163 & 6620 & 39783 & 0.9272 & 0.4816 & 0.4474 \\
X-Net & 8385 & 6300 & 14685 & 0.9731 & 0.7191 & 0.7050 \\
ACE-Net & 11675 & 6991 & 18666 & 0.9658 & 0.6599 & 0.6420 \\
CAAE & 10145 & {\ul 5006} & 15151 & 0.9723 & 0.7262 & 0.7117 \\
SRF & 10824 & 6031 & 16855 & 0.9691 & 0.6935 & 0.6774 \\
CSTN & \textbf{2438} & 5859 & \textbf{8297} & \textbf{0.9848} & \textbf{0.8226} & \textbf{0.8147} \\ \hline
\end{tabular}
\end{table}

\begin{table}[t!]
    \centering
    \caption{Performance Comparison Between Different Change Detection Methods on the California Dataset. The Best Results are in Bold, and the Second-best are Underlined.\label{tab:CM_California}}
\renewcommand\arraystretch{1.2}
\begin{tabular}{c|cccccc}
\hline
Method & FP & FN & OE & OA & F1 & KC \\ \hline
INLPG & 20780 & 9479 & 30259 & 0.9308 & {\ul 0.5353} & {\ul 0.4992} \\
IRG-McS & 7154 & 16557 & 23711 & 0.9458 & 0.4661 & 0.4389 \\
SCCN & 41611 & {\ul 8933} & 50544 & 0.8845 & 0.4156 & 0.3615 \\
PMBCN & 49168 & 10157 & 59325 & 0.8644 & 0.3609 & 0.2997 \\
CAN & 39020 & 13295 & 52315 & 0.8804 & 0.3422 & 0.2840 \\
X-Net & 28596 & 11644 & 40240 & 0.9080 & 0.4314 & 0.3845 \\
ACE-Net & 26248 & 11795 & 38043 & 0.9130 & 0.4427 & 0.3979 \\
CAAE & 27933 & \textbf{7345} & 35278 & 0.9194 & 0.5259 & 0.4885 \\
SRF & {\ul 6468} & 15189 & {\ul 21657} & {\ul 0.9505} & 0.5197 & 0.4947 \\
CSTN & \textbf{5457} & 14051 & \textbf{19508} & \textbf{0.9554} & \textbf{0.5686} & \textbf{0.5460} \\ \hline
\end{tabular}
\end{table}

\begin{figure*}[htbp!]
    \centering
    \includegraphics[width=0.80\textwidth]{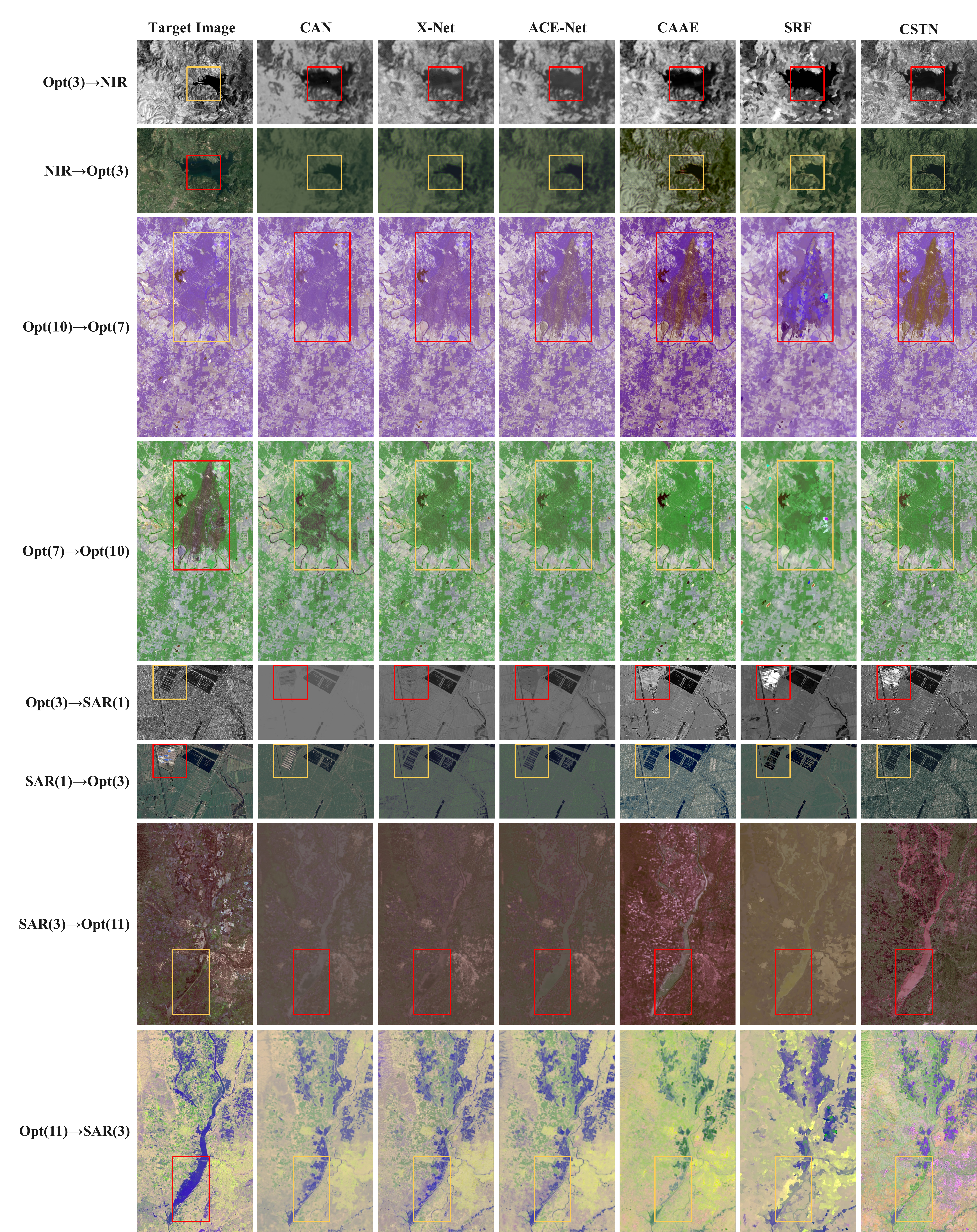}
	\caption{Visual results of image translation across four datasets. From top to bottom, the results correspond to the Sardinia, Texas, Shuguang and California datasets. The rectangular sections highlight the core components of dataset change areas, emphasizing the preservation of content between original and translated images marked in corresponding color-coded boxes, as well as the similarity in stylistic attributes with target images.}
    \label{fig:Translated_Image}
\end{figure*}
\subsection{Discussion}
In this section, we delve into the comparative evaluation of image translation performance and conduct an ablation study to dissect the loss components contributing to the effectiveness of our proposed CSTN model.
\subsubsection{Image Translation Performance}
To substantiate our framework's efficacy, this section presents a comparative evaluation of the image translation output of CSTN against five other image translation-based change detection methods. Visual image translation outputs are shown in Fig. \ref{fig:Translated_Image}, While the corresponding quantitative results are presented in Table \ref{tab:I2I_FID_KID}. CSTN demonstrates strong performance in both visual outputs and quantitative metrics for image translation tasks. Importantly, we note that high-quality image translation does not always correspond directly to improved change detection performance, highlighting a crucial consideration for utilizing image translation in change detection tasks.

As shown in Table \ref{tab:I2I_FID_KID}, CSTN achieves the highest FID and KID scores in most image translation tasks. Although we did not obtain the top scores for the Texas dataset and the Opt(11)→SAR(3) translation in the California dataset, our quantitative results remain in the upper tier. For instance, in the Texas Opt(10)→Opt(7) translation, CSTN achieves FID and KID scores of 62.48 and 4.82, respectively. While CAN outperformed CSTN with scores of 48.59 and 3.80, CSTN's scores were notably closer to optimal performance compared to the worst-performing method, SRF (FID: 205.82, KID: 23.05). Despite CAN's superior quantitative results, its translation in the Texas Opt(10)→Opt(7) direction loses change regions, as depicted in Fig. \ref{fig:Translated_Image} (third row, as highlighted by the red rectangle), whereas CSTN better preserves the original content. This observation underscores CSTN's impressive image translation capability, effectively balancing target domain style with content preservation. This reinforces the effectiveness of our proposed workflow. Although our change detection component does not directly rely on translated images, our model excels in both tasks, showcasing the overall efficacy of our approach.

In addition, among the five image translation-based change detection methods, there is no clear positive correlation between quantitative image translation scores and change detection performance. Taking the Texas dataset as an example, where change regions are prominent, Table \ref{tab:CM_Texas} shows that CAAE achieves the best change detection results with F1 and KC scores of 0.8757 and 0.8593, respectively, among methods like CAN, X-Net, ACE-Net, and SRF. However, Table \ref{tab:I2I_FID_KID} reveals that CAAE's image translation performance ranks second to last in both translation directions. In contrast, CAN, despite poor change detection results, scores higher in image translation quality. CAN's translations heavily reflect the target domain's content, whereas CAAE better preserves the original content, leading to superior change detection capabilities. This discrepancy suggests that current image translation metrics primarily emphasize similarity to the target image without distinguishing between content and style. Consequently, translations heavily influenced by target domain content score higher in quality but may compromise change detection accuracy. Effective change detection through image translation requires preserving original image content as much as possible for direct comparison, despite inevitable transformations during translation. Therefore, CSTN's approach of training the model for image translation and conducting change detection in a style-independent content feature space appears to be a more rational choice.

\begin{table*}[tbp!]
    \centering
    \caption{Image Translation Performance Comparison Between Different Methods on the Four Datasets. The Best Results are in Bold.\label{tab:I2I_FID_KID}}
\renewcommand\arraystretch{1.2}
\begin{tabular}{c|cccccccc}
\hline
Dataset & Domain & Indicator & CAN & X-Net & ACE-Net & CAAE & SRF & CSTN \\ \hline
\multirow{4}{*}{Sardinia} & \multirow{2}{*}{Opt(3)→NIR} & FID↓ & 291.80 & 150.62 & 198.83 & 131.99 & 376.75 & \textbf{109.50} \\
 &  & KID×100↓ & 32.55 & 12.11 & 18.56 & 7.56 & 42.70 & \textbf{4.00} \\ \cline{2-9} 
 & \multirow{2}{*}{NIR→Opt(3)} & FID↓ & 280.37 & 200.91 & 218.91 & 222.34 & 255.25 & \textbf{133.81} \\
 &  & KID×100↓ & 33.72 & 19.23 & 24.18 & 20.60 & 27.79 & \textbf{10.88} \\ \hline
\multirow{4}{*}{Texas} & \multirow{2}{*} {Opt(10)→ Opt(7)} & FID↓ & \textbf{48.59} & 54.30 & 90.39 & 117.53 & 205.82 & 62.48 \\
 &  & KID×100↓ & \textbf{3.80} & 4.51 & 9.19 & 12.36 & 23.05 & 4.82 \\\cline{2-9} 
 & \multirow{2}{*}{Opt(7)→ Opt(10)} & FID↓ & 56.88 & \textbf{51.79} & 84.31 & 102.54 & 200.54 & 65.06 \\
 &  & KID×100↓ & 4.06 & \textbf{4.29} & 8.81 & 10.07 & 22.10 & 5.62 \\ \hline
\multirow{4}{*}{Shuguang} & \multirow{2}{*}{Opt(3)→SAR(1)} & FID↓ & 390.46 & 202.69 & 221.69 & 229.16 & 316.49 & \textbf{189.22} \\
 &  & KID×100↓ & 53.15 & 23.04 & 25.40 & 23.82 & 38.96 & \textbf{18.48} \\ \cline{2-9} 
 & \multirow{2}{*}{SAR(1)→Opt(3)} & FID↓ & 240.06 & 251.51 & 271.04 & 223.04 & 242.02 & \textbf{222.84} \\
 &  & KID×100↓ & 22.93 & 22.77 & 25.68 & 19.89 & 20.33 & \textbf{19.49} \\  \hline
\multirow{4}{*}{California} & \multirow{2}{*}{SAR(3)→Opt(11)} & FID↓ & 225.24 & 185.37 & 193.51 & 219.66 & 266.45 & \textbf{149.10} \\
 &  & KID×100↓ & 21.76 & 16.39 & 16.90 & 19.41 & 25.90 & \textbf{10.46} \\ \cline{2-9} 
 & \multirow{2}{*} {Opt(11)→SAR(3)} & FID↓ & 145.41 & \textbf{99.22} & 140.11 & 191.03 & 178.34 & 111.78 \\
 &  & KID×100↓ & 11.37 & \textbf{6.17} & 11.44 & 18.91 & 14.48 & 7.84 \\\hline
\end{tabular}
\end{table*}

\subsubsection{Ablation Study}
The ablation study aimes to evaluate the impact of each loss component on our model's performance. We selected the California dataset for its extensive and diverse data, which encompasses complex real-world scenarios, making it an ideal platform for this analysis.

The results presented in Table \ref{tab:ablation} underscore the critical importance of each loss component. It is evident from the data that incorporating all loss components leads to optimal model performance, reflected in the lowest OE of 19508, highest OA of 0.9554, and highest KC of 0.5460. This highlights the collaborative effect of these components in achieving peak performance.

Omitting individual components results in a noticeable decline in performance. The reconstruction loss directly optimizes the encoders and decoders by focusing on within-domain image encoding and reconstruction. Excluding it leads to the highest OE and lowest KC, indicating its significant impact on the model's ability to separate content and style effectively. Similarly, the cycle consistency loss and the translation loss enhance the model's capacity through cross-domain translation and reconstruction tasks. By incorporating these more complex workflows, these components significantly boost the model's efficacy. Under the constraints of the above three loss components, the model demonstrates relatively strong feature extraction capabilities. However, the addition of the content code alignment loss further enhances performance in MCD tasks, showing an additional improvement in the model's ability to align content codes across domains. 

In summary, the complexity of the California dataset provided a rigorous testing ground, demonstrating that integrating reconstruction, transformation, cycle consistency and content code alignment losses is crucial for achieving high accuracy and reliability in MCD tasks.

\begin{table}[tbp]
\centering
\caption{Ablation Study on the California Dataset.\label{tab:ablation}}
\renewcommand\arraystretch{1.5}
\begin{tabular}{ccccccc}
\hline
$\mathcal{L} _{recon}$ & $\mathcal{L} _{tans}$ & $\mathcal{L}_{cyc}$ & $\mathcal{L} _{align}$ & OE & OA & KC \\ \hline
 & \usym{1F5F8} & \usym{1F5F8} & \usym{1F5F8} & 51358 & 0.8826 & 0.3485 \\
\usym{1F5F8} &  & \usym{1F5F8} & \usym{1F5F8} & 19942 & 0.9544 & 0.5194 \\
\usym{1F5F8} & \usym{1F5F8} &  & \usym{1F5F8} & 28295 & 0.9353 & 0.5025 \\
\usym{1F5F8} & \usym{1F5F8} & \usym{1F5F8} &  & 19952 & 0.9544 & 0.5433 \\ \hline
\usym{1F5F8} & \usym{1F5F8} & \usym{1F5F8} & \usym{1F5F8} & \textbf{19508} & \textbf{0.9554} & \textbf{0.5460} \\ \hline
\end{tabular}
\end{table}

\section{Conclusion}
\label{Section:Conclusion}
This study introduces a novel unsupervised framework CSTN for MCD in remote sensing, addressing the critical challenge of identifying changes between images captured by different sensors. Our proposed model effectively separates content and style information through dual-domain multi-task workflow. This approach facilitates accurate image reconstruction and transformation without relying on adversarial training. The synergistic integration of reconstruction, translation, cycle consistency, and a novel content code alignment loss has been instrumental in optimizing the model's performance, leading to substantial improvements over existing state-of-the-art methods.

The primary contributions of this work are threefold. First, we have successfully established a style-independent content-comparable feature space, effectively mitigating the challenges posed by variations in sensor types and imaging conditions inherent in multimodal remote sensing data. Second, we have introduced a unified network architecture that seamlessly integrates image translation and change detection tasks, thereby enhancing both performance. Moreover, the simplification of the training process, achieved through the equal weighting of loss components, significantly reduces the complexity of hyperparameter tuning, enhancing the model's practicality and usability. Future work will explore integrating attention mechanisms for enhanced sensitivity and interpretability, and multi-task learning paradigms for addressing diverse imaging conditions and related remote sensing tasks.

\bibliographystyle{IEEEtran}
\bibliography{reference}

\end{document}